\documentclass[%
 amsmath,amssymb,
preprint,%
superscriptaddress,
]{revtex4-1}

\usepackage{graphicx}% Include figure files
\usepackage{dcolumn}% Align table columns on decimal point
\usepackage{bm}% bold math
\usepackage[mathlines]{lineno}% Enable numbering of text and display math
\usepackage[utf8]{inputenc}
\usepackage[T1]{fontenc}
\usepackage{mathptmx}
\usepackage{etoolbox}
\usepackage{url}            % simple URL typesetting
\usepackage{subfiles}
\usepackage{array}
\newcolumntype{P}[1]{>{\centering\arraybackslash}m{#1}}
\usepackage{lineno}
\usepackage{microtype}
\usepackage{amsmath}
\usepackage{amssymb}
\usepackage{mathtools}
\usepackage{txfonts}
\setcitestyle{super}

\begin{document}

\title{Dephasingless laser wakefield acceleration in the bubble regime}
% Force line breaks with \\
\author{Kyle G. Miller}
\email{kmill@lle.rochester.edu}
\affiliation{Laboratory for Laser Energetics, University of Rochester, Rochester, NY 14623-1299, USA}%Lines break automatically or can be forced with \\

\author{Jacob R. Pierce}
% \email{jacobpierce@physics.ucla.edu}
\affiliation{Department of Physics and Astronomy, University of California, Los Angeles, CA 90095, USA}

\author{Manfred V. Ambat}
% \email{mvambat@ur.rochester.edu}
\affiliation{Laboratory for Laser Energetics, University of Rochester, Rochester, NY 14623-1299, USA}

\author{Jessica L. Shaw}
% \email{jshaw05@lle.rochester.edu}
\affiliation{Laboratory for Laser Energetics, University of Rochester, Rochester, NY 14623-1299, USA}

\author{Kale Weichman}
% \email{kweic@lle.rochester.edu}
\affiliation{Laboratory for Laser Energetics, University of Rochester, Rochester, NY 14623-1299, USA}

\author{Warren B. Mori}
% \email{mori@physics.ucla.edu}
\affiliation{Department of Physics and Astronomy, University of California, Los Angeles, CA 90095, USA}
\affiliation{Department of Electrical and Computer Engineering, University of California, Los Angeles, CA 90095, USA}

\author{Dustin H. Froula}
% \email{dfroula@lle.rochester.edu}
\affiliation{Laboratory for Laser Energetics, University of Rochester, Rochester, NY 14623-1299, USA}

\author{John P. Palastro}
% \email{jpal@lle.rochester.edu}
\affiliation{Laboratory for Laser Energetics, University of Rochester, Rochester, NY 14623-1299, USA}

\date{\today}

\begin{abstract} % 200 words
Laser wakefield accelerators (LWFAs) have electric fields that are orders of magnitude larger than those of conventional accelerators, promising an attractive, small-scale alternative for next-generation light sources and lepton colliders. The maximum energy gain in a single-stage LWFA is limited by dephasing, which occurs when the trapped particles outrun the accelerating phase of the wakefield. Here, we demonstrate that a single space--time structured laser pulse can be used for ionization injection and electron acceleration over many dephasing lengths in the bubble regime. Simulations of a dephasingless laser wakefield accelerator driven by a 6.2-J laser pulse show 25~pC of injected charge accelerated over 20~dephasing lengths (1.3~cm) to a maximum energy of 2.1~GeV. The space--time structured laser pulse features an ultrashort, programmable-trajectory focus. Accelerating the focus, reducing the focused spot-size variation, and mitigating unwanted self-focusing stabilize the electron acceleration, which improves beam quality and leads to projected energy gains of 125~GeV in a single, sub-meter stage driven by a 500-J pulse.
\end{abstract}

% \begin{keyword}
% Particle-in-cell (PIC) \sep Laser wakefield acceleration \sep Betatron resonance \sep Direct laser acceleration
% \end{keyword}

\maketitle

\section{Introduction} \label{sec:intro}

In a laser wakefield accelerator (LWFA), the ponderomotive force of an ultrashort laser pulse propagating through plasma displaces electrons and excites a large-amplitude plasma wave.\cite{Tajima1979,Esarey2009}  The fields of the plasma wave can exceed 100~GV/m and are orders of magnitude larger than those of conventional radio-frequency accelerators. The next generation of LWFAs may provide ultra-compact, high-energy colliders and advanced light sources.\cite{Albert2016} To do so, however, these LWFAs will have to address three factors that can limit the maximum energy gain: diffraction, depletion, and dephasing. Of these three, dephasing---the advance of a high-energy electron from the accelerating to decelerating phase of the plasma wave---is typically the most difficult to address.\cite{Durfee1993,Spence2000,Leemans2006,Gonsalves2019,Sprangle1988,Clayton2010,Wang2013,Kim2013,Lu2007} State-of-the-art, single-stage LWFAs operate at low density ($10^{17}$~cm$^{-3}$) to achieve the highest electron energies (${\lesssim}10$~GeV) over a single dephasing length (${\sim}20$~cm).\cite{Gonsalves2019,Schroeder2023,Lu2007}  Acceleration past these energies requires either multiple stages\cite{Steinke2016,Schroeder2023} or some technique to circumvent dephasing.\cite{Yoon2014,Zhang2015,Debus2019,Palastro2020}

\begin{figure}[!htp]
    %\centering
\includegraphics[width=0.45\linewidth]{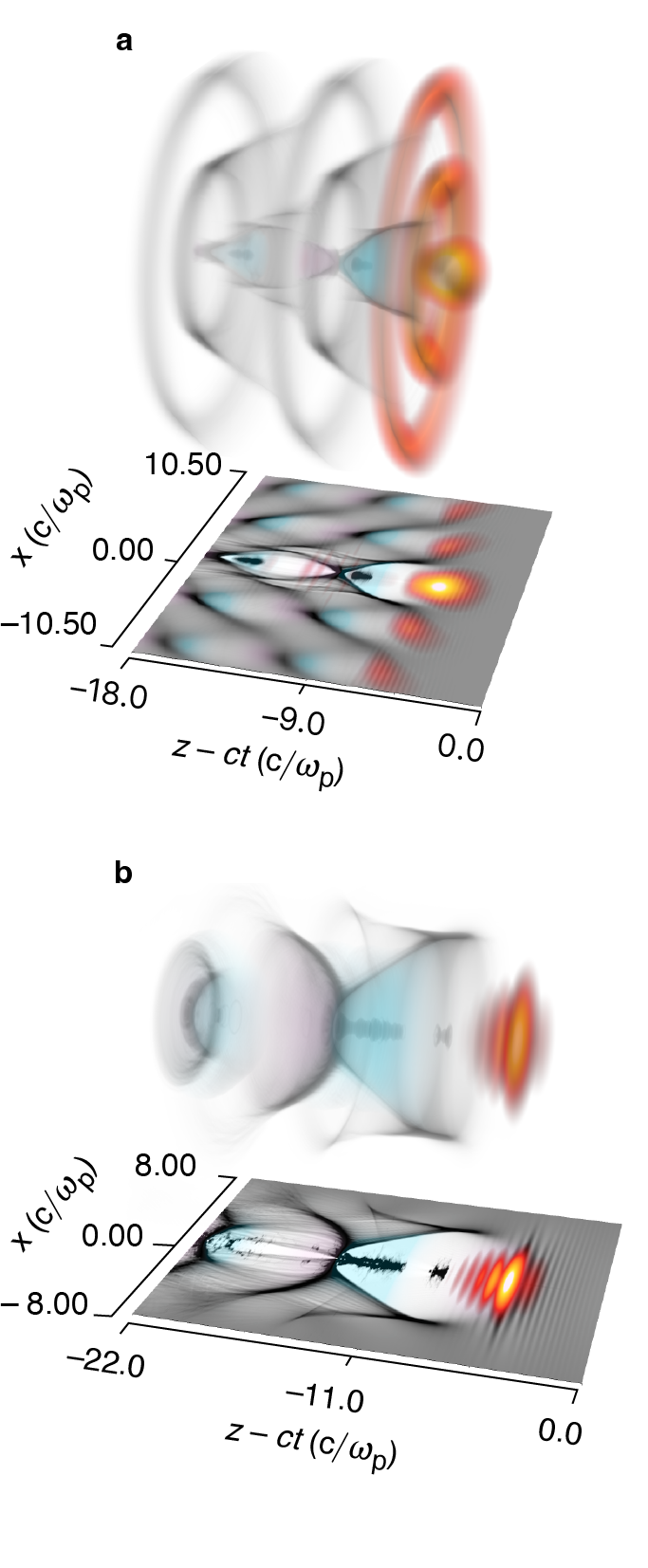}
\caption{\label{fig:DLWFA-LWFA} Comparison of dephasingless and traditional laser wakefield accelerators. (a)~In a dephasingless laser wakefield accelerator, fresh light rays continually come into focus to produce a near-luminal intensity peak and wake, thereby preventing dephasing. (b)~In a traditional laser wakefield accelerator, the trapped electrons eventually outrun the accelerating phase of the wakefield, limiting the maximum energy gain.  Contours of laser intensity (red/yellow), electron density (gray), and accelerating/decelerating wakefield (teal/pink) are shown. The first bubble (white) trails the laser pulse and is devoid of all but the trapped electrons.}
\end{figure}

Spatiotemporal structuring of light can produce laser pulses that feature a programmable-trajectory ``flying focus'' that travels distances far greater than a Rayleigh range while maintaining a near-constant profile.\cite{Sainte-Marie2017,Froula2018,Palastro2020,Jolly2020,Ambat2023} The ultrafast flying focus, in particular, uses an axiparabola to focus different near-field annuli of a laser pulse to different longitudinal locations and the radial group delay imparted by a radial echelon to control the timing of those foci.\cite{Smartsev2019,Oubrerie2022,Palastro2020,Pigeon2023,Ambat2023} The resulting ultrashort intensity peak can be made to travel at the vacuum speed of light inside a plasma, making it ideal for a dephasingless laser wakefield accelerator (DLWFA).\cite{Palastro2020,Caizergues2020}  As fresh light rays come into focus, they continually drive a luminal wake [Fig.~\ref{fig:DLWFA-LWFA}(a)], simultaneously solving the issues of diffraction, depletion, and dephasing present in a traditional LWFA [Fig.~\ref{fig:DLWFA-LWFA}(b)]. This allows DLWFAs to operate at high density ($10^{19}$~cm$^{-3}$), where the accelerating fields are stronger.

The DLWFA concept has been demonstrated in simulations of linear and quasi-linear wakes that used either an external beam or a density downramp to inject electrons into the wake.\cite{Debus2019,Palastro2020,Caizergues2020,Geng2022,Geng2023,Palastro2021}  The first simulations of a DLWFA driven by an ultrafast flying focus showed energy gains of ${>}1$~GeV over ${\sim}1$~cm for externally injected beams.\cite{Palastro2020,Caizergues2020}  Further investigations yielded insight into the field stability\cite{Geng2022} and used a 5-fold, 10-$\muup$m density downramp to inject and accelerate 10~pC of charge to ${\sim}400$~MeV.\cite{Geng2023}  A DLWFA operating in the nonlinear bubble regime,\cite{Lu2007, Froula2009,Mangles2012} where plasma electrons are completely expelled from the path of the laser pulse, could take advantage of even larger accelerating fields. Operation in this regime would allow for self-injection from a uniform plasma or ionization injection, obviating the need for tailored density gradients. Ionization injection, in particular, offers the potential for smaller-emittance beams with lower laser powers.\cite{Chen2006,Pak2010,McGuffey2010,Pollock2011,Xu2014,Zhang2016} Regardless of the injection mechanism, stable propagation of a flying focus has yet to be demonstrated in the bubble regime.

In this work, we demonstrate ionization injection and stable acceleration of electrons in a bubble-regime dephasingless laser wakefield accelerator. Using a single 6.2-J pulse, 25~pC of charge are injected and accelerated over 20~dephasing lengths, or 1.3~cm, to a maximum (average) energy of 2.1 (1.7)~GeV. Structuring the flying-focus pulse to control the motion of the bubble enables the generation of a high-quality electron beam with a 1.8\% average energy spread and 2.2 mm-mrad normalized emittance.  This is done by accelerating the focus to compensate for a changing spot size, masking the inner portion of the axiparabola--echelon pair to reduce the amount of light trapped in the bubble, and positioning the plasma to mitigate unwanted self-focusing. Scaling these results to near-term experiments with 500~J of laser pulse energy suggests that energy gains of 125~GeV over a distance of ${<}1$~m are possible.

\section{Results}

\begin{figure}[!htp]
    %\centering
\includegraphics[width=0.5\linewidth]{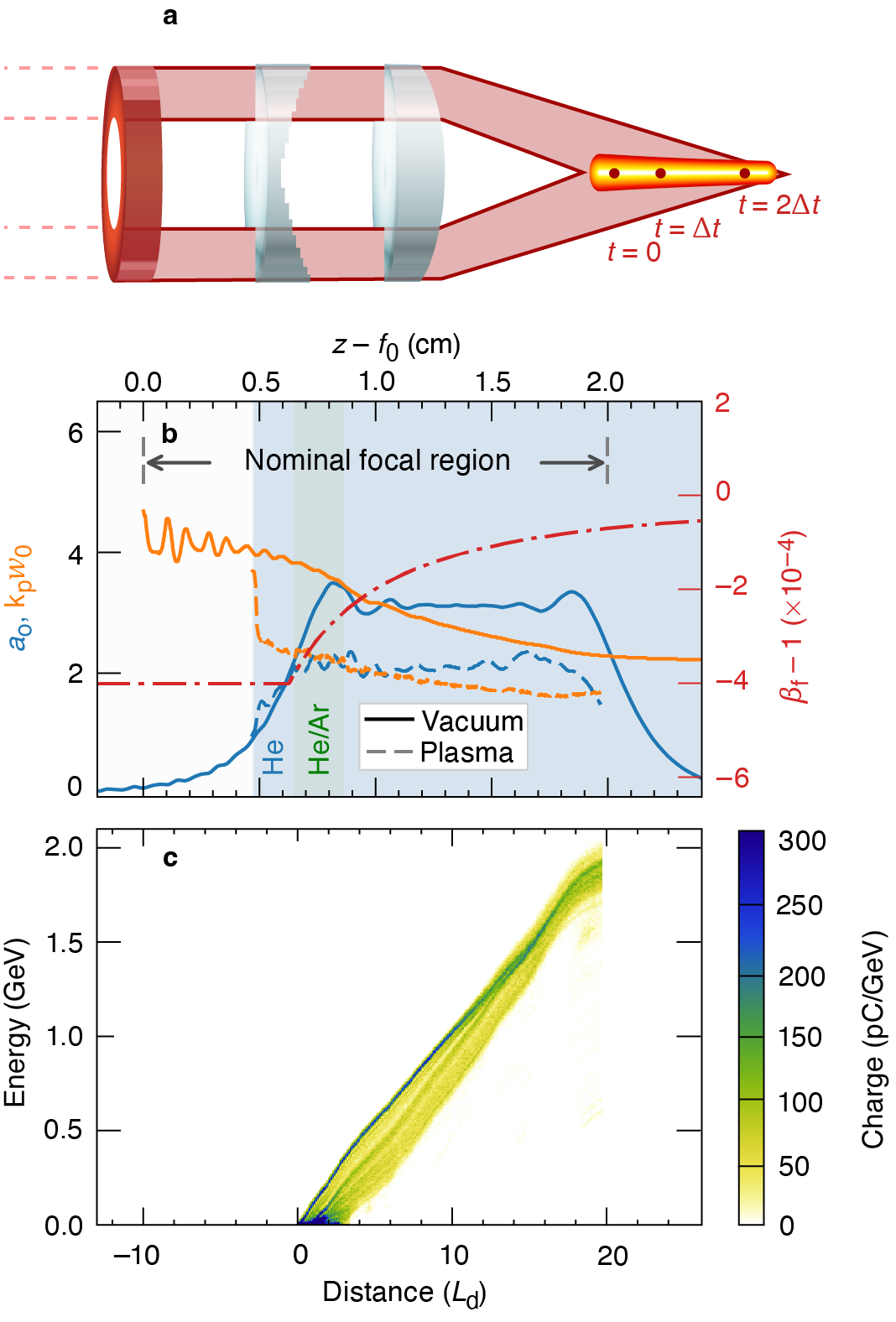}
\caption{\label{fig:main-result} The ultrafast flying focus and electron acceleration in a bubble-regime dephasingless laser wakefield accelerator. (a)~Schematic of the optical configuration for an accelerating focus, including the axiparabola and echelon. For illustrative purposes, the optics are shown in transmission, but experiments would likely be performed in reflection.\cite{Pigeon2023} (b)~The accelerator geometry showing the on-axis amplitude $a_0$ and inner-core spot size $w_0$ of the masked laser pulse---simulated in vacuum (solid) and plasma (dashed)---along with the designed focal velocity in the plasma $\beta_\mathrm{f}$ (dot-dashed). (c)~Energy gain of the ionization-injected electrons in the first bubble. After 20~dephasing lengths, 25~pC of charge was accelerated up to 2.1~GeV. The inset displays a snapshot of the He electron density along with the trapped (Ar) electron density.}
\end{figure}

To demonstrate ionization injection and acceleration in a DLWFA, particle-in-cell simulations were conducted for a flying-focus pulse generated by an axiparabola and a radial echelon [Fig.~\ref{fig:main-result}(a)]. The axiparabola produces an extended focal region, and the echelon imparts a radial group delay that provides control over the trajectory of the focus.  Figure~\ref{fig:main-result}(b) shows a schematic of the nominal focal region, focal velocity, laser amplitude, and spot size produced by these optics along with the plasma density.  Three key modifications to the original DLWFA design enable stable acceleration in the bubble regime: (i)~accelerating the focus to maintain trapping and acceleration of the injected electron beam, (ii)~masking the inner portion of the optics to eliminate laser light where the focused spot size is largest, and (iii)~starting the plasma sufficiently upstream of the peak intensity to reduce unwanted self-focusing.

In the simulations, a flying-focus pulse with wavelength $\lambda_0 = 1.054\,\muup$m, duration $\tau = 15$~fs matched to the plasma density ($\pi/\tau\mathrm \approx \omega_\mathrm{p}$, where $\omega_\mathrm{p}$ is the plasma frequency), and a peak vacuum intensity $I_0 = 1.1\times 10^{19}$~W/cm$^2$ propagated through a preionized He$^\mathrm{2+}$ plasma locally doped with Ar$^\mathrm{8+}$.  The pulse further ionized the argon and drove a nonlinear wake over 1.3~cm (20~dephasing lengths $L_\mathrm{d}$). The freed electrons were injected over 2.2~mm, and the resulting 25-pC beam was accelerated to a maximum (average) energy of 2.1 (1.7)~GeV [Fig.~\ref{fig:main-result}(c)].  Near the end of the focal region, the normalized beam emittance was 2.2~mm-mrad with an average energy spread of 1.8\%.  The laser pulse had 6.2~J of energy, resulting in a laser-to-beam efficiency of 0.7\%.
% 43 mJ in the electron beam
The average accelerating gradient was approximately 1.5~GeV/cm (0.25~GeV/J) in terms of the accelerator length (pulse energy). This compares favorably to the record experimental result for traditional LWFA of 0.4~GeV/cm (0.25~GeV/J).\cite{Gonsalves2019}

Ionization injection requires a sufficiently large laser electric field and bubble radius to trap electrons. Ionized electrons born near the peak of the laser pulse experience a drop in their potential energy and a corresponding increase in their longitudinal momentum as they drift to the rear of the bubble. The electrons are trapped if they move through a change in the wake potential $\psi$ that satisfies\cite{Oz2007,Pak2010}
\begin{equation} \label{eq:trapping}
   \Delta \psi = \psi_\mathrm{f} - \psi_\mathrm{i} \lesssim -1, 
\end{equation}
where $\psi = e(\phi - \beta_\mathrm{w}A_z)/mc^2$, $\beta_\mathrm{w} = \varv_\mathrm{w}/c$ is the normalized wake speed, $\phi$ is the electrostatic potential, and $A_z$ is the longitudinal vector potential. In order to satisfy Eq.~\eqref{eq:trapping}, ionization injection\cite{Pak2010,McGuffey2010,Zhang2016} typically requires laser pulses with amplitudes $a_0 \gtrsim 2$, spot sizes within the plasma $k_\mathrm{p} w_0 \gtrsim 2$, and powers $P/P_\mathrm{c} \gtrsim 0.5$, where $a_0 \approx 8.55 \times 10^{-10} \lambda_0\,(\muup\mathrm{m}) \sqrt{I_0\,(\mathrm{W}/\mathrm{cm}^2)}$ is the normalized vector potential of the laser pulse, $k_\mathrm{p} = \omega_\mathrm{p}/c$, and $P_\mathrm{c}$ is the critical power for relativistic self-focusing.\cite{Sun1987} This contrasts evolving-bubble self-injection, where trapping is typically only observed if $P/P_\mathrm{c} \gtrsim 1$.  \cite{Froula2009,Mangles2012}

Producing a stable wake structure and controlling the focal trajectory enables the trapping, retention, and acceleration of ionized electrons with a single pulse.  Although relativistic self-focusing can be used to guide a conventional laser pulse,\cite{Sprangle1988,Clayton2010,Wang2013,Kim2013} the same process can disrupt the transverse structure of a flying-focus pulse and produce deleterious modulations in the spot size and on-axis intensity.\cite{Geng2022}  These modulations can perturb the electron sheath, change the bubble size and shape, and result in a loss of trapped charge or poor beam quality.  In addition, self-focusing and refraction from the nonlinear plasma structure can cause the on-axis and first off-axis radial maxima of the pulse [Fig.~\ref{fig:DLWFA-LWFA}(a)] to merge. This doubles the power in the radial core of the pulse and further exacerbates the effects of nonlinear propagation.  The simulations performed in this work suggest that a condition for stable propagation of a flying-focus pulse is given by
\begin{equation} \label{eq:power}
    \frac{P_0}{P_\mathrm{c}} \approx \frac{(a_0 k_\mathrm{p} w_0)^2}{32} \lesssim 0.5,
\end{equation}
where $P_0$ is the power integrated out to the first radial minimum of the intensity when the pulse first enters the plasma. To date, stable propagation in a DLWFA has only been demonstrated for $a_0 \leq 1.5$.\cite{Caizergues2020,Geng2022,Geng2023} The remainder of this section describes the design of a stable DLWFA with $a_0 \gtrsim 2.0$ and sufficient power for ionization injection.

% Peak (average) energy is 2.06 (1.72) GeV

A subluminal and accelerating focal trajectory [Fig. \ref{fig:main-result}(b)] prevents the back of the wake from eclipsing the trapped charge and positions the charge in the strongest accelerating field. The radially dependent focal length of the axiparabola, $f(r) = f_0 + (r/R)^2 L$, focuses different near-field radii $r \leq R$ to different longitudinal locations $z=f(r)$ within the focal range $L$. As a result, the inner core of the flying focus pulse has a vacuum spot size $w_\mathrm{v}(z) = f(z) \lambda_0/\pi r(z)$ that decreases along the focal region:
\begin{equation}
    w_\mathrm{v}(z) \approx \frac{\lambda_0 f_0}{\pi R} \sqrt{\frac{L}{z-f_0}},
\end{equation}
where $f_0 \gg L >0$ is assumed. The radius of the bubble is approximately equal to this spot size. For a constant-velocity flying focus, the decreasing spot size causes the rear sheath of the bubble to accelerate and eventually overtake the trapped charge. This can be avoided by programming the focal trajectory so that the back of the bubble moves at the vacuum speed of light:
\begin{equation} \label{eq:betaf}
    \beta_\mathrm{f}(z) = \beta_\mathrm{v}\frac{\varv_\mathrm{g}}{c} = 1 + \alpha \frac{\mathrm{d}w_\mathrm{v}}{\mathrm{d}z} = 1-\alpha \frac{\lambda_0 f_0}{2\pi R}\sqrt{\frac{L}{(z-f_0)^3}},
\end{equation}
where $\beta_\mathrm{f} \equiv \varv_\mathrm{f}/c$ and $\beta_\mathrm{v}$ are the normalized focal velocities in plasma and vacuum, respectively, $\varv_\mathrm{g} = (1 - \omega_\mathrm{p}^2/\omega_0^2)^{1/2}$ is the group velocity of the laser pulse, $\omega_0 = 2\pi c/\lambda_0$, and $\alpha = 0.6$ is a numerically determined factor that accounts for the reduction in spot size due to nonlinear propagation. The focal point accelerates so that $\beta_\mathrm{f}(z)$ asymptotes to unity with increasing distance.

\begin{figure}[!htp]
    %\centering
\includegraphics[width=0.5\linewidth]{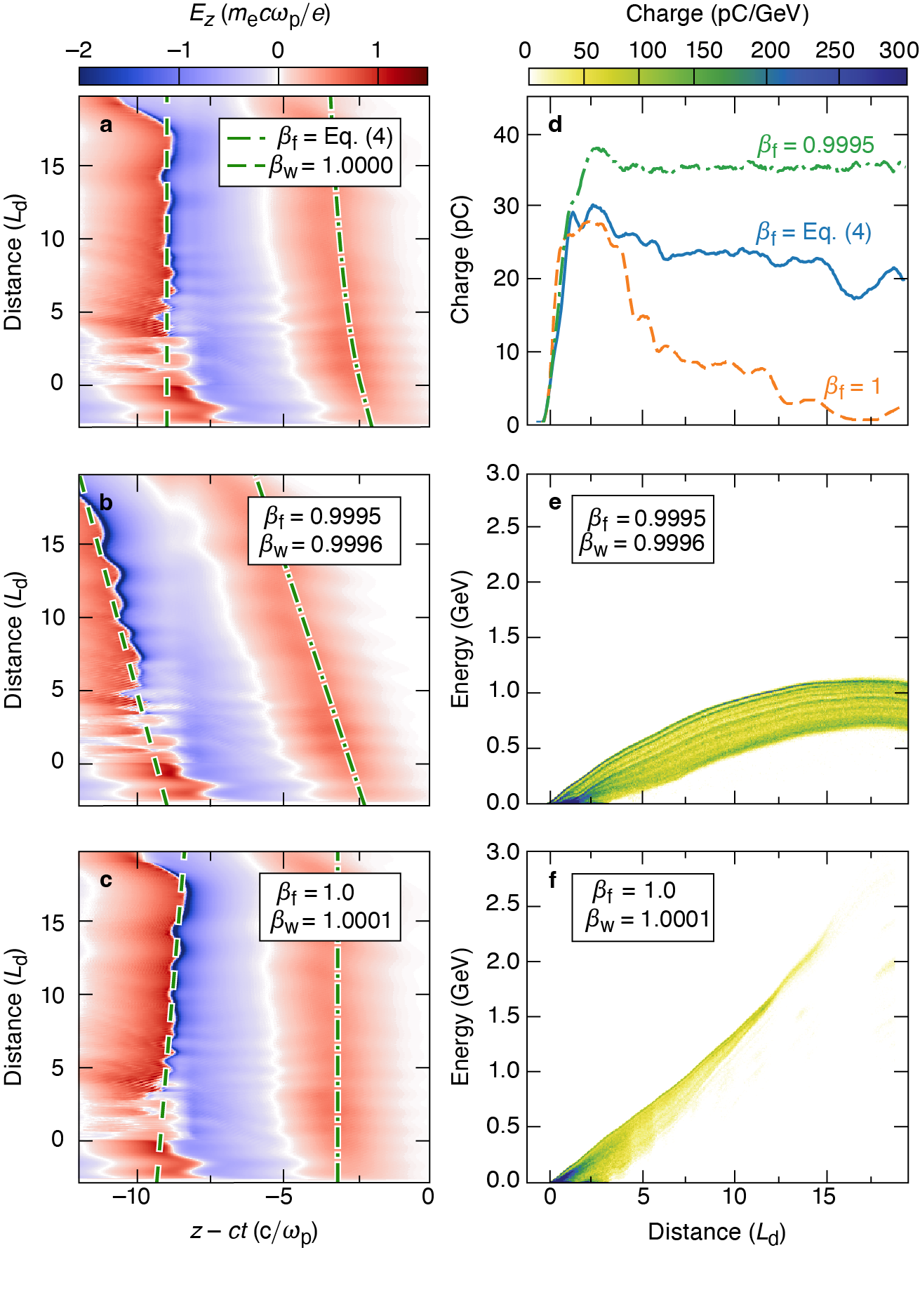}
\caption{\label{fig:velocity-comparison} Dependence of plasma wave and electron beam properties on the focal trajectory. The on-axis, longitudinal electric field of the wake $E_z$ for normalized focal velocities (a)~specified by Eq.~\eqref{eq:betaf}, (b)~set to 0.9995, and (c)~set to 1.0. The resulting normalized velocity of the back of the wake $\beta_\mathrm{w}$ (dashed-dot) and the vacuum speed of light trajectory (dashed) are also shown. (d)~The total trapped charge in the first bubble for cases (a)--(c).  (e) and (f)~Energy gain of the ionization-injected electrons in the first bubble for the cases (b) and (c), respectively. Only for the accelerating focus was the trapped charge both accelerated and maintained over the entire focal region.}
\end{figure}

Accounting for the evolution of the bubble when specifying the focal trajectory [as in Eq.~\eqref{eq:betaf}] prevents dephasing and a loss of trapped charge.  Figure~\ref{fig:velocity-comparison} shows the on-axis, longitudinal electric field of the wake $E_z$ for three different focal trajectories: in (a), the focal velocity from Eq.~\eqref{eq:betaf} produced a luminal wake that is optimal for electron acceleration and maintaining the trapped charge [Fig.~\ref{fig:main-result}(c)]; in (b), a subluminal focus drove a subluminal wake that resulted in dephasing [Fig.~\ref{fig:velocity-comparison}(e)]; and in (c), a luminal focus drove a superluminal wake that overtook and lost the trapped charge [Fig.~\ref{fig:velocity-comparison}(f)]. The total trapped charge in the first bubble for these cases is displayed in Fig.~\ref{fig:velocity-comparison}(d). Only the accelerating focus, as specified by Eq.~\eqref{eq:betaf}, both maintained and accelerated the trapped charge over the entire focal region.

\begin{figure}[!htp]
    %\centering
\includegraphics[width=\linewidth]{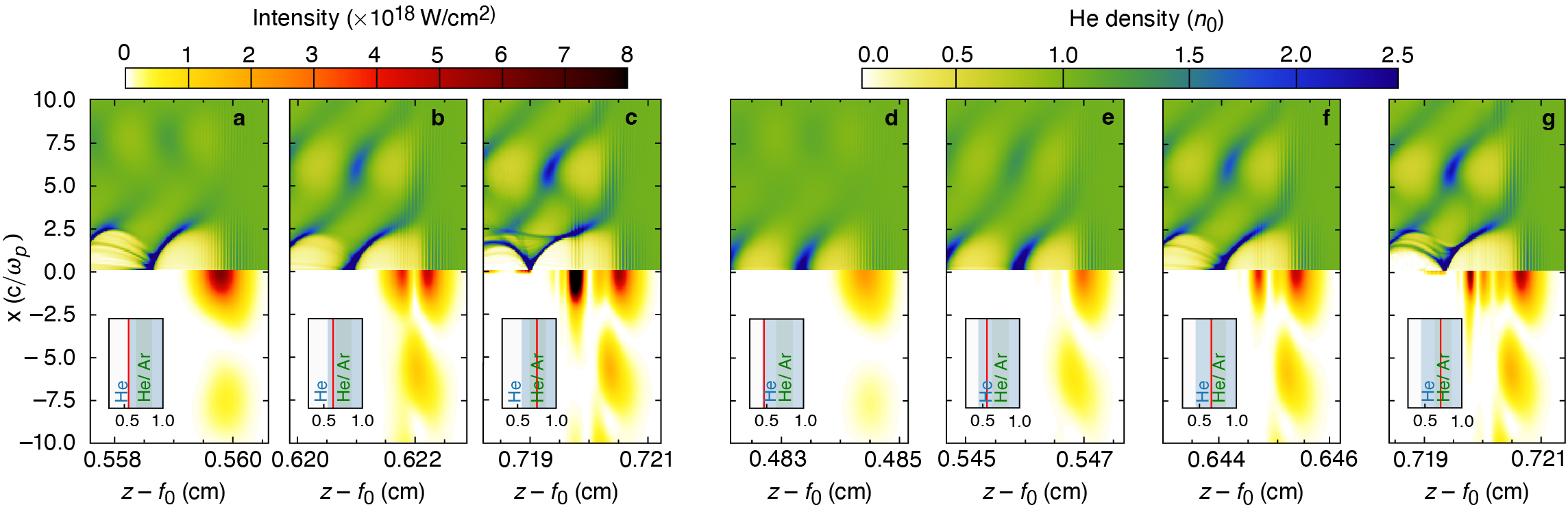}
\caption{\label{fig:snapshots} Laser pulse and bubble evolution in a dephasingless laser wakefield accelerator. Snapshots of the laser intensity (bottom) and plasma density (top) at various distances for the schematic in Fig.~\ref{fig:main-result}(b). In (a)--(c), the plasma begins 0.54~cm into the focal region. In (d)--(g), the plasma begins 0.465~cm into the focal region. Panels (c) and (g) correspond to the same spatial location, but the deformation of the bubble due to trapped light is only observed in (c).}
\end{figure}

Masking an inner portion of the axiparabola and echelon reduces the spot-size variation of the focused pulse and stabilizes its propagation through the focal region. With the laser amplitudes $a_0 > 2$ needed for ionization injection and operation in the bubble regime, stable propagation of the flying-focus pulse requires that $k_\mathrm{p} w_0 < 2$ [Eq.~\eqref{eq:power}]. However, the spot size of an ultrashort flying-focus pulse varies significantly (often by a factor of 4) over the focal region,\cite{Smartsev2019,Geng2022,Oubrerie2022,Ambat2023} making it impossible to satisfy $k_\mathrm{p} w_0 < 2$ everywhere. This can be resolved by eliminating the section of the focal region where the spot size is largest.  For all results here, the axiparabola and echelon optics were masked from $0$ to $0.54R$ to remove the first 30\% of the focal region. The resulting spot size varied by a factor of only ${\sim}1.6$ (${\sim}1.4$) in vacuum (plasma) over the shortened region [Fig.~\ref{fig:main-result}(b)]. 
% Spot size varies from 3.78 to 2.29 in vacuum and 2.29 to 1.61 in plasma

The position of the plasma relative to the focal region also plays a critical role in ensuring stable propagation and in meeting the requirement of Eq.~\eqref{eq:power}. Figure~\ref{fig:snapshots} displays temporal snapshots of the laser pulse envelope and electron density for two plasma configurations. In (a)--(c), the plasma began 0.54~cm into the focal region ($L=2$~cm). The large amplitude of the pulse as it entered the plasma resulted in strong self-focusing and the trapping of an intense sub-pulse that deformed the bubble [see (c)]. In (d)--(g), the plasma began earlier at 0.465~cm into the focal region. Starting the plasma at this location (or even earlier), where the amplitude of the pulse is smaller, mitigates self-focusing and the trapping of light within the bubble. This prevents significant deformation of the bubble, as demonstrated by comparing the two cases at equal distances [cf. (c) and (g)].

\section{Discussion}

The laser-to-beam energy efficiency of 0.7\% quoted in the Results section could be improved upon by increasing the amount of trapped charge or its energy gain. In the simulations, the accelerating field varied longitudinally along the electron beam: the field was stronger near the rear of the bubble and weaker closer to the center [Fig.~\ref{fig:velocity-comparison}(a)]. Loading the wake would produce a flat accelerating field and potentially reduce the final electron energy spread.\cite{Katsouleas1987,Tzoufras2008,Dalichaouch2021} The amount of trapped charge could be increased by extending the argon-doped region, enlarging the normalized spot size $k_\mathrm{p}w_0$, or using a density downramp.  Fine-tuning the focal velocity to position the beam closer to the back of the bubble could increase the energy gain [see Fig.~\ref{fig:velocity-comparison}(f), where electrons were accelerated to nearly 3~GeV with a faster focal velocity]. Experimentally, the beam charge and energy could be optimized in real time by adjusting the focal trajectory using a deformable mirror and spatial light modulator pair instead of an echelon.\cite{Sun2015,Li2020} 

The efficiency could also be increased by structuring the transverse profile of the laser pulse. The radial intensity profile incident on the axiparabola could be shaped to reduce the value of $a_0$ after the ionization-injection region. This would place the accelerator in a more-linear regime and increase the efficiency at the cost of a longer accelerator.\cite{Tzoufras2012}

Nonlinear propagation and transverse structures in the plasma density reduce the laser pulse amplitude and spot size relative to their vacuum values [Fig.~\ref{fig:main-result}(b)]. The axiparabola maps different annuli in the near field to different longitudinal locations in the far field. The resulting interference produces a radial intensity profile with concentric maxima [Fig.~\ref{fig:DLWFA-LWFA}(a)]. When the flying-focus pulse has sufficient amplitude, these maxima can ponderomotively drive ring-like plasma waves that channel, trap, and deplete some of the laser light. For the case considered here, this reduced the on-axis value of $a_0$ from 3 in vacuum to ${\sim}2.2$ in the plasma.  In addition, relativistic self-focusing and channeling in the ring-like structures caused a rapid and sustained decrease in the spot size of each maximum. A flying-focus pulse with an amplitude $a_0 \ll 1$ did not produce these structures and propagated identically to the vacuum case, but with the focal velocity reduced by a factor of $\varv_\mathrm{g}/c$.

The stability of the accelerating structure is expected to improve with accelerator length.  The on-axis intensity modulations visible in the vacuum $a_0$ shown in Fig.~\ref{fig:main-result}(b) restrict the rear positioning of the electron beam in the wake. The modulations cause the bubble to oscillate, which can result in the rear sheath overtaking and detrapping the electron beam. The amplitude of the intensity modulations decreases for longer focal regions (see Methods section), meaning that a more-optimal beam placement should be possible for larger accelerator lengths.

Radial masking of the optics enhances the efficiency and stability of a DLWFA but is not strictly required to accelerate over many dephasing lengths.  When the optics were left unmasked for the case shown in Fig.~\ref{fig:main-result}, the laser pulse and bubble exhibited highly nonlinear evolution that was isolated to the beginning of the focal region. The large spot size at the beginning of the focal region resulted in substantial self-focusing followed by stochastic trapping, acceleration, and the loss of ionized electrons. Any self-focused light that was trapped within the bubble [as in Fig.~\ref{fig:snapshots}(c)] propagated slightly slower than the speed of light and was eventually left behind.  Thus, stable propagation, ionization injection, and acceleration still occurred farther into the focal region, albeit with lower charge and energy gains than in the masked case.

More experimentally feasible alternatives to a fully preionized plasma could also be investigated. The simulations presented here assumed a preionized plasma with a transverse and longitudinal extent of 2.6~mm and 1.5~cm, respectively. For a meter-long accelerator ($L=1$~m) using an axiparabola with the same f-number $f_\# = f_0/2R = 7$, the plasma would have to be preionized over a 15~cm diameter, which may be experimentally infeasible.  A DLWFA could instead be designed so that the flying-focus pulse itself ionizes the plasma.  However, this would require adjustments to the focal velocity and may change the nonlinear plasma response, which will require further investigation via simulation. Finally, other realizations of a flying focus or use of structured light may allow for additional optimization.\cite{Pierce2023,Yessenov2022,Simpson2022,Li2020a}.

In conclusion, ionization injection and stable acceleration over 20~dephasing lengths in a bubble-regime DLWFA has been demonstrated. A stable accelerating structure was attained by (i)~prescribing an accelerating focal trajectory to compensate for the changing spot size produced by the axiparabola, (ii)~masking the interior of the optics to reduce the variation of the spot size within the plasma, and (iii)~beginning the plasma farther into the focal region to mitigate self-focusing. With the same accelerating gradient, a 500-J laser pulse driving a DLWFA over ${\sim}80$~cm could produce an energy gain of 125~GeV. Further optimization of the focal trajectory could result in even higher acceleration gradients and efficiencies.

% TODO: Run matched cases for a0 = 3 and a0 = 2, then make some claims about the stability of the a0 and w0

\section{Methods} % <3000 words
\label{sec:methods}

\subsection{Dephasing length}

In a traditional laser wakefield accelerator, the laser pulse travels slower than the vacuum speed of light. Near-luminal electrons trapped in the wake can advance relative to the pulse and outrun the accelerating phase of the wakefield, a process known as dephasing.  In the bubble regime, the accelerating field changes sign near the center of the bubble.\cite{Lu2006} Thus, the electrons reach their maximum energy after advancing approximately one bubble radius relative to the laser pulse, i.e., upon moving from the back to the center of the bubble. The distance over which this occurs---the dephasing length---depends on the velocity of the front edge of the laser pulse and the bubble radius.\cite{Lu2007} Specifically, 
\begin{equation}
L_\mathrm{d} = \frac{2}{3} \frac{\omega_0^2}{\omega_\mathrm{p}^2} w_0,
\end{equation}
where it has been assumed that the bubble radius is approximately equal to the spot size. In Ref.~\onlinecite{Lu2007}, a matching condition $k_\mathrm{p} w_0 = 2\sqrt{a_0}$ was determined that leads to stable propagation for a traditional LWFA in the bubble regime. If this condition is met, the dephasing length is then given by $k_\mathrm{p} L_\mathrm{d} = \frac{4}{3} (\omega_0^2/\omega_\mathrm{p}^2) \sqrt{a_0}$.

When comparing the DLWFA [Fig.~\ref{fig:main-result}(b)] to a traditional LWFA, various choices can be made in determining an equivalent dephasing length.  The vacuum values of $a_0 = 3$ and $k_\mathrm{p} w_0 = 3.5$ could be used (which are approximately matched), yielding a dephasing length of $L_\mathrm{d} \approx 331 k_\mathrm{p}^{-1} \approx 0.665\,$mm.  If the value of $a_0 = 2.2$ in the plasma is used instead, then $L_\mathrm{d} \approx 284 k_\mathrm{p}^{-1} \approx 0.569\,$mm assuming a matched spot size.  Alternatively, the value of $k_\mathrm{p} w_0 = 2.2$ in the plasma could be used to obtain $L_\mathrm{d} \approx 210 k_\mathrm{p}^{-1} \approx 0.422\,$mm. All comparisons between the DLWFA and a traditional LWFA made in this work use the first and most-conservative choice, $L_\mathrm{d} \approx 0.665\,$mm, which corresponds to the simulation shown in Fig.~\ref{fig:DLWFA-LWFA}(b).

\subsection{Design of the axiparabola, radial group delay, and radial chirp}

The initial laser fields used in the PIC simulation were obtained by propagating the laser pulse from the flying-focus optical assembly to the start of the simulation domain using a frequency-domain Fresnel integral.\cite{Palastro2018} The optical assembly applied three modifications to the laser pulse: (i)~the phase from an axiparabola to focus each annulus of the pulse to a different longitudinal location; (ii)~the radial group delay from an echelon to control the focal trajectory; and (iii)~a chirp that varied with radius to preemptively invert group-velocity dispersion in the plasma. The axiparabola essentially uses spherical aberration to extend the focal region.\cite{Smartsev2019,Geng2022} Here, a positive focal range ($L>0$) was used so that the largest spot size occurs at the beginning of the focal region to better facilitate ionization injection. The echelon consisted of concentric rings of half-wavelength ($\lambda_0/2$) depth and variable widths determined by the desired focal trajectory.\cite{Palastro2020,Ambat2023} The radial chirp can be introduced by applying a variable-thickness coating to the surface of the echelon. For more-adaptive control over the focal trajectory, the echelon can be replaced by a deformable mirror and spatial light modulator.\cite{Ambat2023}

The lineouts of vacuum $a_0$ and spot size $w_0$ shown in Fig.~\ref{fig:main-result}(b) were computed by evaluating the Fresnel integral at an initial point in the far field, then using the unidirectional pulse propagation
equation\cite{Kolesik2002,Kolesik2004,Couairon2011} to model the laser propagation.  For all results shown, the modeled axiparabola had a radius $R=5$~cm, a nominal focal length $f_0=70$~cm, and a nominal focal range of $L=2$~cm.  The laser pulse had a wavelength of $\lambda_0 = 1.054\,\muup$m and a Gaussian temporal profile with an intensity FWHM of 15~fs.

\subsection{Particle-in-cell simulations}

All PIC simulations were performed using the quasi-3D geometry of \textsc{Osiris},\cite{Fonseca2002a,Fonseca2013,Davidson2015} where modes 0 and 1 were retained in the azimuthal expansion.  A customized field solver that mitigates errors from the numerical dispersion relation and the time-staggering of the electromagnetic fields was employed.\cite{Li2021,Miller2023} As a result, no extraneous numerical corrections had to be made to the focal trajectory of the pulse, as has been done in prior simulations of dephasingless laser wakefield accelerators.\cite{Caizergues2020,Geng2022,Geng2023}

For the simulation pictured in Fig.~\ref{fig:main-result}, the preionized background plasma was simulated with 32 particles per cell ($2\times2\times8$) out to a radius of $40\,c/\omega_\mathrm{p}$ and 8 particles per cell ($1\times1\times8$) thereafter. The 9--14 levels of unionized argon electrons were simulated with a possible 8 particles per cell per level out to a radius of $15\,c/\omega_\mathrm{p}$. The preionized plasma had an 80-$\muup$m upramp (results were insensitive to the size of all upramps) followed by a uniform density of $7\times10^{18}\,$cm$^{-3}$. In the ionization-injection region, this density was obtained via a 90\%~He$^{2+}$/10\%~Ar$^{8+}$ mix (resulting in a 69\%/31\% respective contribution to the preionized electron background).  The grid had $4106\times6570$ cells in $z\times r$, with 30~points per laser wavelength and 10~points per plasma period, respectively. The time step was $0.0102\,\omega_\mathrm{p}^{-1}$. Altogether, the simulation used a $143\,\muup\mathrm{m}\times1.32\,$mm box and a total propagation distance (time) of  1.51~cm (50.2~ps).

\subsection{On-axis intensity modulations of the flying-focus pulse}

The on-axis electric field of a laser pulse focused by an axiparabola can be expressed as
\begin{equation}
    E(z,\omega_0) = \frac{\omega_0}{icz}e^{-iA^2/4B}\int_0^R \exp{\left[ iB \left(r'^2+\frac{A}{2B} \right)^2 \right]} r'\,dr',
\end{equation}
where $A = \omega_0/2c (z^{-1}-f_0^{-1})$ and $B = \omega_0 L/4cf_0^2R^2$.  Integrating this expression and taking the squared norm (see Appendix C of Ref.~\onlinecite{Ambat2023}) results in an expression that exhibits modulations of scale length
\begin{equation}
    L_\mathrm{m} \approx \frac{2f_0}{R} \left( \frac{cL}{\omega_0} \right)^{1/2}.
\end{equation}
The modulation length increases with the focal length and focal range and decreases with the axiparabola radius. Suppressing the amplitude of the modulations requires a focal range $L \gg L_\mathrm{m} \sqrt{2\pi}$ or
\begin{equation}
    L \gg 16 f_\#^2 \lambda_0,
\end{equation}
where $f_\#$ is the f-number. In the simulations presented here, $L = 20\,\mathrm{mm}$ and $16 f_\#^2 \lambda_0 = 0.8\,\mathrm{mm}$.  For the same $f_\#$, longer acceleration lengths would reduce the amplitude of the modulations even further. 

\begin{acknowledgments}
This report was prepared as an account of work sponsored by an agency of the U.S. Government. Neither the U.S. Government nor any agency thereof, nor any of their employees, makes any warranty, express or implied, or assumes any legal liability or responsibility for the accuracy, completeness, or usefulness of any information, apparatus, product, or process disclosed, or represents that its use would not infringe privately owned rights. Reference herein to any specific commercial product, process, or service by trade name, trademark, manufacturer, or otherwise does not necessarily constitute or imply its endorsement, recommendation, or favoring by the U.S. Government or any agency thereof. The views and opinions of authors expressed herein do not necessarily state or reflect those of the U.S. Government or any agency thereof.

This material is based upon work supported by the Office of Fusion Energy Sciences under Award Numbers DE-SC00215057 and DE-SC0010064, the Department of Energy National Nuclear Security Administration under Award Numbers DE-NA0003856 and DE-NA0004131, the National Science Foundation under Award Number 2108970, the University of Rochester, and the New York State Energy Research and Development Authority.  Simulations were performed at NERSC under m4372.
\end{acknowledgments}

\section*{Data Availability Statement}

The data that support the findings of this study are available from the corresponding author upon reasonable request.

\bibliography{references-manual-export}% Produces the bibliography via BibTeX.

\begin{thebibliography}{59}%
\makeatletter
\providecommand \@ifxundefined [1]{%
 \@ifx{#1\undefined}
}%
\providecommand \@ifnum [1]{%
 \ifnum #1\expandafter \@firstoftwo
 \else \expandafter \@secondoftwo
 \fi
}%
\providecommand \@ifx [1]{%
 \ifx #1\expandafter \@firstoftwo
 \else \expandafter \@secondoftwo
 \fi
}%
\providecommand \natexlab [1]{#1}%
\providecommand \enquote  [1]{``#1''}%
\providecommand \bibnamefont  [1]{#1}%
\providecommand \bibfnamefont [1]{#1}%
\providecommand \citenamefont [1]{#1}%
\providecommand \href@noop [0]{\@secondoftwo}%
\providecommand \href [0]{\begingroup \@sanitize@url \@href}%
\providecommand \@href[1]{\@@startlink{#1}\@@href}%
\providecommand \@@href[1]{\endgroup#1\@@endlink}%
\providecommand \@sanitize@url [0]{\catcode `\\12\catcode `\$12\catcode
  `\&12\catcode `\#12\catcode `\^12\catcode `\_12\catcode `\%12\relax}%
\providecommand \@@startlink[1]{}%
\providecommand \@@endlink[0]{}%
\providecommand \url  [0]{\begingroup\@sanitize@url \@url }%
\providecommand \@url [1]{\endgroup\@href {#1}{\urlprefix }}%
\providecommand \urlprefix  [0]{URL }%
\providecommand \Eprint [0]{\href }%
\providecommand \doibase [0]{http://dx.doi.org/}%
\providecommand \selectlanguage [0]{\@gobble}%
\providecommand \bibinfo  [0]{\@secondoftwo}%
\providecommand \bibfield  [0]{\@secondoftwo}%
\providecommand \translation [1]{[#1]}%
\providecommand \BibitemOpen [0]{}%
\providecommand \bibitemStop [0]{}%
\providecommand \bibitemNoStop [0]{.\EOS\space}%
\providecommand \EOS [0]{\spacefactor3000\relax}%
\providecommand \BibitemShut  [1]{\csname bibitem#1\endcsname}%
\let\auto@bib@innerbib\@empty
%</preamble>
\bibitem [{\citenamefont {Tajima}\ and\ \citenamefont
  {Dawson}(1979)}]{Tajima1979}%
  \BibitemOpen
  \bibfield  {author} {\bibinfo {author} {\bibfnamefont {T.}~\bibnamefont
  {Tajima}}\ and\ \bibinfo {author} {\bibfnamefont {J.~M.}\ \bibnamefont
  {Dawson}},\ }\href {\doibase 10.1103/PhysRevLett.43.267} {\bibfield
  {journal} {\bibinfo  {journal} {Physical Review Letters}\ }\textbf {\bibinfo
  {volume} {43}},\ \bibinfo {pages} {267} (\bibinfo {year} {1979})}\BibitemShut
  {NoStop}%
\bibitem [{\citenamefont {Esarey}\ \emph {et~al.}(2009)\citenamefont {Esarey},
  \citenamefont {Schroeder},\ and\ \citenamefont {Leemans}}]{Esarey2009}%
  \BibitemOpen
  \bibfield  {author} {\bibinfo {author} {\bibfnamefont {E.}~\bibnamefont
  {Esarey}}, \bibinfo {author} {\bibfnamefont {C.~B.}\ \bibnamefont
  {Schroeder}}, \ and\ \bibinfo {author} {\bibfnamefont {W.~P.}\ \bibnamefont
  {Leemans}},\ }\href {\doibase 10.1103/RevModPhys.81.1229} {\bibfield
  {journal} {\bibinfo  {journal} {Reviews of Modern Physics}\ }\textbf
  {\bibinfo {volume} {81}},\ \bibinfo {pages} {1229} (\bibinfo {year}
  {2009})}\BibitemShut {NoStop}%
\bibitem [{\citenamefont {Albert}\ and\ \citenamefont
  {Thomas}(2016)}]{Albert2016}%
  \BibitemOpen
  \bibfield  {author} {\bibinfo {author} {\bibfnamefont {F.}~\bibnamefont
  {Albert}}\ and\ \bibinfo {author} {\bibfnamefont {A.~G.~R.}\ \bibnamefont
  {Thomas}},\ }\href {\doibase 10.1088/0741-3335/58/10/103001} {\bibfield
  {journal} {\bibinfo  {journal} {Plasma Physics and Controlled Fusion}\
  }\textbf {\bibinfo {volume} {58}},\ \bibinfo {pages} {103001} (\bibinfo
  {year} {2016})}\BibitemShut {NoStop}%
\bibitem [{\citenamefont {Durfee}\ and\ \citenamefont
  {Milchberg}(1993)}]{Durfee1993}%
  \BibitemOpen
  \bibfield  {author} {\bibinfo {author} {\bibfnamefont {C.~G.}\ \bibnamefont
  {Durfee}}\ and\ \bibinfo {author} {\bibfnamefont {H.~M.}\ \bibnamefont
  {Milchberg}},\ }\href {\doibase 10.1103/PhysRevLett.71.2409} {\bibfield
  {journal} {\bibinfo  {journal} {Physical Review Letters}\ }\textbf {\bibinfo
  {volume} {71}},\ \bibinfo {pages} {2409} (\bibinfo {year}
  {1993})}\BibitemShut {NoStop}%
\bibitem [{\citenamefont {Spence}\ and\ \citenamefont
  {Hooker}(2000)}]{Spence2000}%
  \BibitemOpen
  \bibfield  {author} {\bibinfo {author} {\bibfnamefont {D.~J.}\ \bibnamefont
  {Spence}}\ and\ \bibinfo {author} {\bibfnamefont {S.~M.}\ \bibnamefont
  {Hooker}},\ }\href {\doibase 10.1103/PhysRevE.63.015401} {\bibfield
  {journal} {\bibinfo  {journal} {Physical Review E}\ }\textbf {\bibinfo
  {volume} {63}},\ \bibinfo {pages} {015401} (\bibinfo {year}
  {2000})}\BibitemShut {NoStop}%
\bibitem [{\citenamefont {Leemans}\ \emph {et~al.}(2006)\citenamefont
  {Leemans}, \citenamefont {Nagler}, \citenamefont {Gonsalves}, \citenamefont
  {Tóth}, \citenamefont {Nakamura}, \citenamefont {Geddes}, \citenamefont
  {Esarey}, \citenamefont {Schroeder},\ and\ \citenamefont
  {Hooker}}]{Leemans2006}%
  \BibitemOpen
  \bibfield  {author} {\bibinfo {author} {\bibfnamefont {W.~P.}\ \bibnamefont
  {Leemans}}, \bibinfo {author} {\bibfnamefont {B.}~\bibnamefont {Nagler}},
  \bibinfo {author} {\bibfnamefont {A.~J.}\ \bibnamefont {Gonsalves}}, \bibinfo
  {author} {\bibfnamefont {C.}~\bibnamefont {Tóth}}, \bibinfo {author}
  {\bibfnamefont {K.}~\bibnamefont {Nakamura}}, \bibinfo {author}
  {\bibfnamefont {C.~G.~R.}\ \bibnamefont {Geddes}}, \bibinfo {author}
  {\bibfnamefont {E.}~\bibnamefont {Esarey}}, \bibinfo {author} {\bibfnamefont
  {C.~B.}\ \bibnamefont {Schroeder}}, \ and\ \bibinfo {author} {\bibfnamefont
  {S.~M.}\ \bibnamefont {Hooker}},\ }\href {\doibase 10.1038/nphys418}
  {\bibfield  {journal} {\bibinfo  {journal} {Nature Physics}\ }\textbf
  {\bibinfo {volume} {2}},\ \bibinfo {pages} {696} (\bibinfo {year}
  {2006})}\BibitemShut {NoStop}%
\bibitem [{\citenamefont {Gonsalves}\ \emph {et~al.}(2019)\citenamefont
  {Gonsalves}, \citenamefont {Nakamura}, \citenamefont {Daniels}, \citenamefont
  {Benedetti}, \citenamefont {Pieronek}, \citenamefont {de~Raadt},
  \citenamefont {Steinke}, \citenamefont {Bin}, \citenamefont {Bulanov},
  \citenamefont {van Tilborg}, \citenamefont {Geddes}, \citenamefont
  {Schroeder}, \citenamefont {Tóth}, \citenamefont {Esarey}, \citenamefont
  {Swanson}, \citenamefont {Fan-Chiang}, \citenamefont {Bagdasarov},
  \citenamefont {Bobrova}, \citenamefont {Gasilov}, \citenamefont {Korn},
  \citenamefont {Sasorov},\ and\ \citenamefont {Leemans}}]{Gonsalves2019}%
  \BibitemOpen
  \bibfield  {author} {\bibinfo {author} {\bibfnamefont {A.}~\bibnamefont
  {Gonsalves}}, \bibinfo {author} {\bibfnamefont {K.}~\bibnamefont {Nakamura}},
  \bibinfo {author} {\bibfnamefont {J.}~\bibnamefont {Daniels}}, \bibinfo
  {author} {\bibfnamefont {C.}~\bibnamefont {Benedetti}}, \bibinfo {author}
  {\bibfnamefont {C.}~\bibnamefont {Pieronek}}, \bibinfo {author}
  {\bibfnamefont {T.}~\bibnamefont {de~Raadt}}, \bibinfo {author}
  {\bibfnamefont {S.}~\bibnamefont {Steinke}}, \bibinfo {author} {\bibfnamefont
  {J.}~\bibnamefont {Bin}}, \bibinfo {author} {\bibfnamefont {S.}~\bibnamefont
  {Bulanov}}, \bibinfo {author} {\bibfnamefont {J.}~\bibnamefont {van
  Tilborg}}, \bibinfo {author} {\bibfnamefont {C.}~\bibnamefont {Geddes}},
  \bibinfo {author} {\bibfnamefont {C.}~\bibnamefont {Schroeder}}, \bibinfo
  {author} {\bibfnamefont {C.}~\bibnamefont {Tóth}}, \bibinfo {author}
  {\bibfnamefont {E.}~\bibnamefont {Esarey}}, \bibinfo {author} {\bibfnamefont
  {K.}~\bibnamefont {Swanson}}, \bibinfo {author} {\bibfnamefont
  {L.}~\bibnamefont {Fan-Chiang}}, \bibinfo {author} {\bibfnamefont
  {G.}~\bibnamefont {Bagdasarov}}, \bibinfo {author} {\bibfnamefont
  {N.}~\bibnamefont {Bobrova}}, \bibinfo {author} {\bibfnamefont
  {V.}~\bibnamefont {Gasilov}}, \bibinfo {author} {\bibfnamefont
  {G.}~\bibnamefont {Korn}}, \bibinfo {author} {\bibfnamefont {P.}~\bibnamefont
  {Sasorov}}, \ and\ \bibinfo {author} {\bibfnamefont {W.}~\bibnamefont
  {Leemans}},\ }\href {\doibase 10.1103/PhysRevLett.122.084801} {\bibfield
  {journal} {\bibinfo  {journal} {Physical Review Letters}\ }\textbf {\bibinfo
  {volume} {122}},\ \bibinfo {pages} {084801} (\bibinfo {year}
  {2019})}\BibitemShut {NoStop}%
\bibitem [{\citenamefont {Sprangle}\ \emph {et~al.}(1988)\citenamefont
  {Sprangle}, \citenamefont {Esarey}, \citenamefont {Ting},\ and\ \citenamefont
  {Joyce}}]{Sprangle1988}%
  \BibitemOpen
  \bibfield  {author} {\bibinfo {author} {\bibfnamefont {P.}~\bibnamefont
  {Sprangle}}, \bibinfo {author} {\bibfnamefont {E.}~\bibnamefont {Esarey}},
  \bibinfo {author} {\bibfnamefont {A.}~\bibnamefont {Ting}}, \ and\ \bibinfo
  {author} {\bibfnamefont {G.}~\bibnamefont {Joyce}},\ }\href {\doibase
  10.1063/1.100300} {\bibfield  {journal} {\bibinfo  {journal} {Applied Physics
  Letters}\ }\textbf {\bibinfo {volume} {53}},\ \bibinfo {pages} {2146}
  (\bibinfo {year} {1988})}\BibitemShut {NoStop}%
\bibitem [{\citenamefont {Clayton}\ \emph {et~al.}(2010)\citenamefont
  {Clayton}, \citenamefont {Ralph}, \citenamefont {Albert}, \citenamefont
  {Fonseca}, \citenamefont {Glenzer}, \citenamefont {Joshi}, \citenamefont
  {Lu}, \citenamefont {Marsh}, \citenamefont {Martins}, \citenamefont {Mori},
  \citenamefont {Pak}, \citenamefont {Tsung}, \citenamefont {Pollock},
  \citenamefont {Ross}, \citenamefont {Silva},\ and\ \citenamefont
  {Froula}}]{Clayton2010}%
  \BibitemOpen
  \bibfield  {author} {\bibinfo {author} {\bibfnamefont {C.~E.}\ \bibnamefont
  {Clayton}}, \bibinfo {author} {\bibfnamefont {J.~E.}\ \bibnamefont {Ralph}},
  \bibinfo {author} {\bibfnamefont {F.}~\bibnamefont {Albert}}, \bibinfo
  {author} {\bibfnamefont {R.~A.}\ \bibnamefont {Fonseca}}, \bibinfo {author}
  {\bibfnamefont {S.~H.}\ \bibnamefont {Glenzer}}, \bibinfo {author}
  {\bibfnamefont {C.}~\bibnamefont {Joshi}}, \bibinfo {author} {\bibfnamefont
  {W.}~\bibnamefont {Lu}}, \bibinfo {author} {\bibfnamefont {K.~A.}\
  \bibnamefont {Marsh}}, \bibinfo {author} {\bibfnamefont {S.~F.}\ \bibnamefont
  {Martins}}, \bibinfo {author} {\bibfnamefont {W.~B.}\ \bibnamefont {Mori}},
  \bibinfo {author} {\bibfnamefont {A.}~\bibnamefont {Pak}}, \bibinfo {author}
  {\bibfnamefont {F.~S.}\ \bibnamefont {Tsung}}, \bibinfo {author}
  {\bibfnamefont {B.~B.}\ \bibnamefont {Pollock}}, \bibinfo {author}
  {\bibfnamefont {J.~S.}\ \bibnamefont {Ross}}, \bibinfo {author}
  {\bibfnamefont {L.~O.}\ \bibnamefont {Silva}}, \ and\ \bibinfo {author}
  {\bibfnamefont {D.~H.}\ \bibnamefont {Froula}},\ }\href {\doibase
  10.1103/PhysRevLett.105.105003} {\bibfield  {journal} {\bibinfo  {journal}
  {Physical Review Letters}\ }\textbf {\bibinfo {volume} {105}},\ \bibinfo
  {pages} {105003} (\bibinfo {year} {2010})}\BibitemShut {NoStop}%
\bibitem [{\citenamefont {Wang}\ \emph {et~al.}(2013)\citenamefont {Wang},
  \citenamefont {Zgadzaj}, \citenamefont {Fazel}, \citenamefont {Li},
  \citenamefont {Yi}, \citenamefont {Zhang}, \citenamefont {Henderson},
  \citenamefont {Chang}, \citenamefont {Korzekwa}, \citenamefont {Tsai},
  \citenamefont {Pai}, \citenamefont {Quevedo}, \citenamefont {Dyer},
  \citenamefont {Gaul}, \citenamefont {Martinez}, \citenamefont {Bernstein},
  \citenamefont {Borger}, \citenamefont {Spinks}, \citenamefont {Donovan},
  \citenamefont {Khudik}, \citenamefont {Shvets}, \citenamefont {Ditmire},\
  and\ \citenamefont {Downer}}]{Wang2013}%
  \BibitemOpen
  \bibfield  {author} {\bibinfo {author} {\bibfnamefont {X.}~\bibnamefont
  {Wang}}, \bibinfo {author} {\bibfnamefont {R.}~\bibnamefont {Zgadzaj}},
  \bibinfo {author} {\bibfnamefont {N.}~\bibnamefont {Fazel}}, \bibinfo
  {author} {\bibfnamefont {Z.}~\bibnamefont {Li}}, \bibinfo {author}
  {\bibfnamefont {S.~A.}\ \bibnamefont {Yi}}, \bibinfo {author} {\bibfnamefont
  {X.}~\bibnamefont {Zhang}}, \bibinfo {author} {\bibfnamefont
  {W.}~\bibnamefont {Henderson}}, \bibinfo {author} {\bibfnamefont {Y.~Y.}\
  \bibnamefont {Chang}}, \bibinfo {author} {\bibfnamefont {R.}~\bibnamefont
  {Korzekwa}}, \bibinfo {author} {\bibfnamefont {H.~E.}\ \bibnamefont {Tsai}},
  \bibinfo {author} {\bibfnamefont {C.~H.}\ \bibnamefont {Pai}}, \bibinfo
  {author} {\bibfnamefont {H.}~\bibnamefont {Quevedo}}, \bibinfo {author}
  {\bibfnamefont {G.}~\bibnamefont {Dyer}}, \bibinfo {author} {\bibfnamefont
  {E.}~\bibnamefont {Gaul}}, \bibinfo {author} {\bibfnamefont {M.}~\bibnamefont
  {Martinez}}, \bibinfo {author} {\bibfnamefont {A.~C.}\ \bibnamefont
  {Bernstein}}, \bibinfo {author} {\bibfnamefont {T.}~\bibnamefont {Borger}},
  \bibinfo {author} {\bibfnamefont {M.}~\bibnamefont {Spinks}}, \bibinfo
  {author} {\bibfnamefont {M.}~\bibnamefont {Donovan}}, \bibinfo {author}
  {\bibfnamefont {V.}~\bibnamefont {Khudik}}, \bibinfo {author} {\bibfnamefont
  {G.}~\bibnamefont {Shvets}}, \bibinfo {author} {\bibfnamefont
  {T.}~\bibnamefont {Ditmire}}, \ and\ \bibinfo {author} {\bibfnamefont
  {M.~C.}\ \bibnamefont {Downer}},\ }\href {\doibase 10.1038/ncomms2988}
  {\bibfield  {journal} {\bibinfo  {journal} {Nature Communications}\ }\textbf
  {\bibinfo {volume} {4}} (\bibinfo {year} {2013}),\
  10.1038/ncomms2988}\BibitemShut {NoStop}%
\bibitem [{\citenamefont {Kim}\ \emph {et~al.}(2013)\citenamefont {Kim},
  \citenamefont {Pae}, \citenamefont {Cha}, \citenamefont {Kim}, \citenamefont
  {Yu}, \citenamefont {Sung}, \citenamefont {Lee}, \citenamefont {Jeong},\ and\
  \citenamefont {Lee}}]{Kim2013}%
  \BibitemOpen
  \bibfield  {author} {\bibinfo {author} {\bibfnamefont {H.~T.}\ \bibnamefont
  {Kim}}, \bibinfo {author} {\bibfnamefont {K.~H.}\ \bibnamefont {Pae}},
  \bibinfo {author} {\bibfnamefont {H.~J.}\ \bibnamefont {Cha}}, \bibinfo
  {author} {\bibfnamefont {I.~J.}\ \bibnamefont {Kim}}, \bibinfo {author}
  {\bibfnamefont {T.~J.}\ \bibnamefont {Yu}}, \bibinfo {author} {\bibfnamefont
  {J.~H.}\ \bibnamefont {Sung}}, \bibinfo {author} {\bibfnamefont {S.~K.}\
  \bibnamefont {Lee}}, \bibinfo {author} {\bibfnamefont {T.~M.}\ \bibnamefont
  {Jeong}}, \ and\ \bibinfo {author} {\bibfnamefont {J.}~\bibnamefont {Lee}},\
  }\href {\doibase 10.1103/PhysRevLett.111.165002} {\bibfield  {journal}
  {\bibinfo  {journal} {Physical Review Letters}\ }\textbf {\bibinfo {volume}
  {111}} (\bibinfo {year} {2013}),\ 10.1103/PhysRevLett.111.165002}\BibitemShut
  {NoStop}%
\bibitem [{\citenamefont {Lu}\ \emph {et~al.}(2007)\citenamefont {Lu},
  \citenamefont {Tzoufras}, \citenamefont {Joshi}, \citenamefont {Tsung},
  \citenamefont {Mori}, \citenamefont {Vieira}, \citenamefont {Fonseca},\ and\
  \citenamefont {Silva}}]{Lu2007}%
  \BibitemOpen
  \bibfield  {author} {\bibinfo {author} {\bibfnamefont {W.}~\bibnamefont
  {Lu}}, \bibinfo {author} {\bibfnamefont {M.}~\bibnamefont {Tzoufras}},
  \bibinfo {author} {\bibfnamefont {C.}~\bibnamefont {Joshi}}, \bibinfo
  {author} {\bibfnamefont {F.~S.}\ \bibnamefont {Tsung}}, \bibinfo {author}
  {\bibfnamefont {W.~B.}\ \bibnamefont {Mori}}, \bibinfo {author}
  {\bibfnamefont {J.}~\bibnamefont {Vieira}}, \bibinfo {author} {\bibfnamefont
  {R.~A.}\ \bibnamefont {Fonseca}}, \ and\ \bibinfo {author} {\bibfnamefont
  {L.~O.}\ \bibnamefont {Silva}},\ }\href {\doibase
  10.1103/PhysRevSTAB.10.061301} {\bibfield  {journal} {\bibinfo  {journal}
  {Physical Review Special Topics - Accelerators and Beams}\ }\textbf {\bibinfo
  {volume} {10}},\ \bibinfo {pages} {061301} (\bibinfo {year}
  {2007})}\BibitemShut {NoStop}%
\bibitem [{\citenamefont {Schroeder}\ \emph {et~al.}(2023)\citenamefont
  {Schroeder}, \citenamefont {Albert}, \citenamefont {Benedetti}, \citenamefont
  {Bromage}, \citenamefont {Bruhwiler}, \citenamefont {Bulanov}, \citenamefont
  {Campbell}, \citenamefont {Cook}, \citenamefont {Cros}, \citenamefont
  {Downer}, \citenamefont {Esarey}, \citenamefont {Froula}, \citenamefont
  {Fuchs}, \citenamefont {Geddes}, \citenamefont {Gessner}, \citenamefont
  {Gonsalves}, \citenamefont {Hogan}, \citenamefont {Hooker}, \citenamefont
  {Huebl}, \citenamefont {Jing}, \citenamefont {Joshi}, \citenamefont
  {Krushelnick}, \citenamefont {Leemans}, \citenamefont {Lehe}, \citenamefont
  {Maier}, \citenamefont {Milchberg}, \citenamefont {Mori}, \citenamefont
  {Nakamura}, \citenamefont {Osterhoff}, \citenamefont {Palastro},
  \citenamefont {Palmer}, \citenamefont {Põder}, \citenamefont {Power},
  \citenamefont {Shadwick}, \citenamefont {Terzani}, \citenamefont {Thévenet},
  \citenamefont {Thomas}, \citenamefont {van Tilborg}, \citenamefont {Turner},
  \citenamefont {Vafaei-Najafabadi}, \citenamefont {Vay}, \citenamefont
  {Zhou},\ and\ \citenamefont {Zuegel}}]{Schroeder2023}%
  \BibitemOpen
  \bibfield  {author} {\bibinfo {author} {\bibfnamefont {C.}~\bibnamefont
  {Schroeder}}, \bibinfo {author} {\bibfnamefont {F.}~\bibnamefont {Albert}},
  \bibinfo {author} {\bibfnamefont {C.}~\bibnamefont {Benedetti}}, \bibinfo
  {author} {\bibfnamefont {J.}~\bibnamefont {Bromage}}, \bibinfo {author}
  {\bibfnamefont {D.}~\bibnamefont {Bruhwiler}}, \bibinfo {author}
  {\bibfnamefont {S.}~\bibnamefont {Bulanov}}, \bibinfo {author} {\bibfnamefont
  {E.}~\bibnamefont {Campbell}}, \bibinfo {author} {\bibfnamefont
  {N.}~\bibnamefont {Cook}}, \bibinfo {author} {\bibfnamefont {B.}~\bibnamefont
  {Cros}}, \bibinfo {author} {\bibfnamefont {M.}~\bibnamefont {Downer}},
  \bibinfo {author} {\bibfnamefont {E.}~\bibnamefont {Esarey}}, \bibinfo
  {author} {\bibfnamefont {D.}~\bibnamefont {Froula}}, \bibinfo {author}
  {\bibfnamefont {M.}~\bibnamefont {Fuchs}}, \bibinfo {author} {\bibfnamefont
  {C.}~\bibnamefont {Geddes}}, \bibinfo {author} {\bibfnamefont
  {S.}~\bibnamefont {Gessner}}, \bibinfo {author} {\bibfnamefont
  {A.}~\bibnamefont {Gonsalves}}, \bibinfo {author} {\bibfnamefont
  {M.}~\bibnamefont {Hogan}}, \bibinfo {author} {\bibfnamefont
  {S.}~\bibnamefont {Hooker}}, \bibinfo {author} {\bibfnamefont
  {A.}~\bibnamefont {Huebl}}, \bibinfo {author} {\bibfnamefont
  {C.}~\bibnamefont {Jing}}, \bibinfo {author} {\bibfnamefont {C.}~\bibnamefont
  {Joshi}}, \bibinfo {author} {\bibfnamefont {K.}~\bibnamefont {Krushelnick}},
  \bibinfo {author} {\bibfnamefont {W.}~\bibnamefont {Leemans}}, \bibinfo
  {author} {\bibfnamefont {R.}~\bibnamefont {Lehe}}, \bibinfo {author}
  {\bibfnamefont {A.}~\bibnamefont {Maier}}, \bibinfo {author} {\bibfnamefont
  {H.}~\bibnamefont {Milchberg}}, \bibinfo {author} {\bibfnamefont
  {W.}~\bibnamefont {Mori}}, \bibinfo {author} {\bibfnamefont {K.}~\bibnamefont
  {Nakamura}}, \bibinfo {author} {\bibfnamefont {J.}~\bibnamefont {Osterhoff}},
  \bibinfo {author} {\bibfnamefont {J.}~\bibnamefont {Palastro}}, \bibinfo
  {author} {\bibfnamefont {M.}~\bibnamefont {Palmer}}, \bibinfo {author}
  {\bibfnamefont {K.}~\bibnamefont {Põder}}, \bibinfo {author} {\bibfnamefont
  {J.}~\bibnamefont {Power}}, \bibinfo {author} {\bibfnamefont
  {B.}~\bibnamefont {Shadwick}}, \bibinfo {author} {\bibfnamefont
  {D.}~\bibnamefont {Terzani}}, \bibinfo {author} {\bibfnamefont
  {M.}~\bibnamefont {Thévenet}}, \bibinfo {author} {\bibfnamefont
  {A.}~\bibnamefont {Thomas}}, \bibinfo {author} {\bibfnamefont
  {J.}~\bibnamefont {van Tilborg}}, \bibinfo {author} {\bibfnamefont
  {M.}~\bibnamefont {Turner}}, \bibinfo {author} {\bibfnamefont
  {N.}~\bibnamefont {Vafaei-Najafabadi}}, \bibinfo {author} {\bibfnamefont
  {J.-L.}\ \bibnamefont {Vay}}, \bibinfo {author} {\bibfnamefont
  {T.}~\bibnamefont {Zhou}}, \ and\ \bibinfo {author} {\bibfnamefont
  {J.}~\bibnamefont {Zuegel}},\ }\href {\doibase
  10.1088/1748-0221/18/06/T06001} {\bibfield  {journal} {\bibinfo  {journal}
  {Journal of Instrumentation}\ }\textbf {\bibinfo {volume} {18}},\ \bibinfo
  {pages} {T06001} (\bibinfo {year} {2023})}\BibitemShut {NoStop}%
\bibitem [{\citenamefont {Steinke}\ \emph {et~al.}(2016)\citenamefont
  {Steinke}, \citenamefont {van Tilborg}, \citenamefont {Benedetti},
  \citenamefont {Geddes}, \citenamefont {Schroeder}, \citenamefont {Daniels},
  \citenamefont {Swanson}, \citenamefont {Gonsalves}, \citenamefont {Nakamura},
  \citenamefont {Matlis}, \citenamefont {Shaw}, \citenamefont {Esarey},\ and\
  \citenamefont {Leemans}}]{Steinke2016}%
  \BibitemOpen
  \bibfield  {author} {\bibinfo {author} {\bibfnamefont {S.}~\bibnamefont
  {Steinke}}, \bibinfo {author} {\bibfnamefont {J.}~\bibnamefont {van
  Tilborg}}, \bibinfo {author} {\bibfnamefont {C.}~\bibnamefont {Benedetti}},
  \bibinfo {author} {\bibfnamefont {C.~G.~R.}\ \bibnamefont {Geddes}}, \bibinfo
  {author} {\bibfnamefont {C.~B.}\ \bibnamefont {Schroeder}}, \bibinfo {author}
  {\bibfnamefont {J.}~\bibnamefont {Daniels}}, \bibinfo {author} {\bibfnamefont
  {K.~K.}\ \bibnamefont {Swanson}}, \bibinfo {author} {\bibfnamefont {A.~J.}\
  \bibnamefont {Gonsalves}}, \bibinfo {author} {\bibfnamefont {K.}~\bibnamefont
  {Nakamura}}, \bibinfo {author} {\bibfnamefont {N.~H.}\ \bibnamefont
  {Matlis}}, \bibinfo {author} {\bibfnamefont {B.~H.}\ \bibnamefont {Shaw}},
  \bibinfo {author} {\bibfnamefont {E.}~\bibnamefont {Esarey}}, \ and\ \bibinfo
  {author} {\bibfnamefont {W.~P.}\ \bibnamefont {Leemans}},\ }\href {\doibase
  10.1038/nature16525} {\bibfield  {journal} {\bibinfo  {journal} {Nature}\
  }\textbf {\bibinfo {volume} {530}},\ \bibinfo {pages} {190} (\bibinfo {year}
  {2016})}\BibitemShut {NoStop}%
\bibitem [{\citenamefont {Yoon}\ \emph {et~al.}(2014)\citenamefont {Yoon},
  \citenamefont {Palastro},\ and\ \citenamefont {Milchberg}}]{Yoon2014}%
  \BibitemOpen
  \bibfield  {author} {\bibinfo {author} {\bibfnamefont {S.}~\bibnamefont
  {Yoon}}, \bibinfo {author} {\bibfnamefont {J.}~\bibnamefont {Palastro}}, \
  and\ \bibinfo {author} {\bibfnamefont {H.}~\bibnamefont {Milchberg}},\ }\href
  {\doibase 10.1103/PhysRevLett.112.134803} {\bibfield  {journal} {\bibinfo
  {journal} {Physical Review Letters}\ }\textbf {\bibinfo {volume} {112}},\
  \bibinfo {pages} {134803} (\bibinfo {year} {2014})}\BibitemShut {NoStop}%
\bibitem [{\citenamefont {Zhang}\ \emph {et~al.}(2015)\citenamefont {Zhang},
  \citenamefont {Khudik},\ and\ \citenamefont {Shvets}}]{Zhang2015}%
  \BibitemOpen
  \bibfield  {author} {\bibinfo {author} {\bibfnamefont {X.}~\bibnamefont
  {Zhang}}, \bibinfo {author} {\bibfnamefont {V.~N.}\ \bibnamefont {Khudik}}, \
  and\ \bibinfo {author} {\bibfnamefont {G.}~\bibnamefont {Shvets}},\ }\href
  {\doibase 10.1103/PhysRevLett.114.184801} {\bibfield  {journal} {\bibinfo
  {journal} {Physical Review Letters}\ }\textbf {\bibinfo {volume} {114}},\
  \bibinfo {pages} {184801} (\bibinfo {year} {2015})}\BibitemShut {NoStop}%
\bibitem [{\citenamefont {Debus}\ \emph {et~al.}(2019)\citenamefont {Debus},
  \citenamefont {Pausch}, \citenamefont {Huebl}, \citenamefont {Steiniger},
  \citenamefont {Widera}, \citenamefont {Cowan}, \citenamefont {Schramm},\ and\
  \citenamefont {Bussmann}}]{Debus2019}%
  \BibitemOpen
  \bibfield  {author} {\bibinfo {author} {\bibfnamefont {A.}~\bibnamefont
  {Debus}}, \bibinfo {author} {\bibfnamefont {R.}~\bibnamefont {Pausch}},
  \bibinfo {author} {\bibfnamefont {A.}~\bibnamefont {Huebl}}, \bibinfo
  {author} {\bibfnamefont {K.}~\bibnamefont {Steiniger}}, \bibinfo {author}
  {\bibfnamefont {R.}~\bibnamefont {Widera}}, \bibinfo {author} {\bibfnamefont
  {T.~E.}\ \bibnamefont {Cowan}}, \bibinfo {author} {\bibfnamefont
  {U.}~\bibnamefont {Schramm}}, \ and\ \bibinfo {author} {\bibfnamefont
  {M.}~\bibnamefont {Bussmann}},\ }\href {\doibase 10.1103/PhysRevX.9.031044}
  {\bibfield  {journal} {\bibinfo  {journal} {Physical Review X}\ }\textbf
  {\bibinfo {volume} {9}},\ \bibinfo {pages} {031044} (\bibinfo {year}
  {2019})}\BibitemShut {NoStop}%
\bibitem [{\citenamefont {Palastro}\ \emph {et~al.}(2020)\citenamefont
  {Palastro}, \citenamefont {Shaw}, \citenamefont {Franke}, \citenamefont
  {Ramsey}, \citenamefont {Simpson},\ and\ \citenamefont
  {Froula}}]{Palastro2020}%
  \BibitemOpen
  \bibfield  {author} {\bibinfo {author} {\bibfnamefont {J.}~\bibnamefont
  {Palastro}}, \bibinfo {author} {\bibfnamefont {J.}~\bibnamefont {Shaw}},
  \bibinfo {author} {\bibfnamefont {P.}~\bibnamefont {Franke}}, \bibinfo
  {author} {\bibfnamefont {D.}~\bibnamefont {Ramsey}}, \bibinfo {author}
  {\bibfnamefont {T.}~\bibnamefont {Simpson}}, \ and\ \bibinfo {author}
  {\bibfnamefont {D.}~\bibnamefont {Froula}},\ }\href {\doibase
  10.1103/PhysRevLett.124.134802} {\bibfield  {journal} {\bibinfo  {journal}
  {Physical Review Letters}\ }\textbf {\bibinfo {volume} {124}},\ \bibinfo
  {pages} {134802} (\bibinfo {year} {2020})}\BibitemShut {NoStop}%
\bibitem [{\citenamefont {Sainte-Marie}\ \emph {et~al.}(2017)\citenamefont
  {Sainte-Marie}, \citenamefont {Gobert},\ and\ \citenamefont
  {Quéré}}]{Sainte-Marie2017}%
  \BibitemOpen
  \bibfield  {author} {\bibinfo {author} {\bibfnamefont {A.}~\bibnamefont
  {Sainte-Marie}}, \bibinfo {author} {\bibfnamefont {O.}~\bibnamefont
  {Gobert}}, \ and\ \bibinfo {author} {\bibfnamefont {F.}~\bibnamefont
  {Quéré}},\ }\href {\doibase 10.1364/OPTICA.4.001298} {\bibfield  {journal}
  {\bibinfo  {journal} {Optica}\ }\textbf {\bibinfo {volume} {4}},\ \bibinfo
  {pages} {1298} (\bibinfo {year} {2017})}\BibitemShut {NoStop}%
\bibitem [{\citenamefont {Froula}\ \emph {et~al.}(2018)\citenamefont {Froula},
  \citenamefont {Turnbull}, \citenamefont {Davies}, \citenamefont {Kessler},
  \citenamefont {Haberberger}, \citenamefont {Palastro}, \citenamefont {Bahk},
  \citenamefont {Begishev}, \citenamefont {Boni}, \citenamefont {Bucht},
  \citenamefont {Katz},\ and\ \citenamefont {Shaw}}]{Froula2018}%
  \BibitemOpen
  \bibfield  {author} {\bibinfo {author} {\bibfnamefont {D.~H.}\ \bibnamefont
  {Froula}}, \bibinfo {author} {\bibfnamefont {D.}~\bibnamefont {Turnbull}},
  \bibinfo {author} {\bibfnamefont {A.~S.}\ \bibnamefont {Davies}}, \bibinfo
  {author} {\bibfnamefont {T.~J.}\ \bibnamefont {Kessler}}, \bibinfo {author}
  {\bibfnamefont {D.}~\bibnamefont {Haberberger}}, \bibinfo {author}
  {\bibfnamefont {J.~P.}\ \bibnamefont {Palastro}}, \bibinfo {author}
  {\bibfnamefont {S.-W.}\ \bibnamefont {Bahk}}, \bibinfo {author}
  {\bibfnamefont {I.~A.}\ \bibnamefont {Begishev}}, \bibinfo {author}
  {\bibfnamefont {R.}~\bibnamefont {Boni}}, \bibinfo {author} {\bibfnamefont
  {S.}~\bibnamefont {Bucht}}, \bibinfo {author} {\bibfnamefont
  {J.}~\bibnamefont {Katz}}, \ and\ \bibinfo {author} {\bibfnamefont {J.~L.}\
  \bibnamefont {Shaw}},\ }\href {\doibase 10.1038/s41566-018-0121-8} {\bibfield
   {journal} {\bibinfo  {journal} {Nature Photonics}\ }\textbf {\bibinfo
  {volume} {12}},\ \bibinfo {pages} {262} (\bibinfo {year} {2018})}\BibitemShut
  {NoStop}%
\bibitem [{\citenamefont {Jolly}\ \emph {et~al.}(2020)\citenamefont {Jolly},
  \citenamefont {Gobert}, \citenamefont {Jeandet},\ and\ \citenamefont
  {Quéré}}]{Jolly2020}%
  \BibitemOpen
  \bibfield  {author} {\bibinfo {author} {\bibfnamefont {S.~W.}\ \bibnamefont
  {Jolly}}, \bibinfo {author} {\bibfnamefont {O.}~\bibnamefont {Gobert}},
  \bibinfo {author} {\bibfnamefont {A.}~\bibnamefont {Jeandet}}, \ and\
  \bibinfo {author} {\bibfnamefont {F.}~\bibnamefont {Quéré}},\ }\href
  {\doibase 10.1364/OE.384512} {\bibfield  {journal} {\bibinfo  {journal}
  {Optics Express}\ }\textbf {\bibinfo {volume} {28}},\ \bibinfo {pages} {4888}
  (\bibinfo {year} {2020})}\BibitemShut {NoStop}%
\bibitem [{\citenamefont {Ambat}\ \emph {et~al.}(2023)\citenamefont {Ambat},
  \citenamefont {Shaw}, \citenamefont {Pigeon}, \citenamefont {Miller},
  \citenamefont {Simpson}, \citenamefont {Froula},\ and\ \citenamefont
  {Palastro}}]{Ambat2023}%
  \BibitemOpen
  \bibfield  {author} {\bibinfo {author} {\bibfnamefont {M.~V.}\ \bibnamefont
  {Ambat}}, \bibinfo {author} {\bibfnamefont {J.~L.}\ \bibnamefont {Shaw}},
  \bibinfo {author} {\bibfnamefont {J.~J.}\ \bibnamefont {Pigeon}}, \bibinfo
  {author} {\bibfnamefont {K.~G.}\ \bibnamefont {Miller}}, \bibinfo {author}
  {\bibfnamefont {T.~T.}\ \bibnamefont {Simpson}}, \bibinfo {author}
  {\bibfnamefont {D.~H.}\ \bibnamefont {Froula}}, \ and\ \bibinfo {author}
  {\bibfnamefont {J.~P.}\ \bibnamefont {Palastro}},\ }\href
  {http://arxiv.org/abs/2307.05313} {\bibfield  {journal} {\bibinfo  {journal}
  {Optics Express}\ } (\bibinfo {year} {2023})},\ \bibinfo {note}
  {accepted}\BibitemShut {NoStop}%
\bibitem [{\citenamefont {Smartsev}\ \emph {et~al.}(2019)\citenamefont
  {Smartsev}, \citenamefont {Caizergues}, \citenamefont {Oubrerie},
  \citenamefont {Gautier}, \citenamefont {Goddet}, \citenamefont {Tafzi},
  \citenamefont {Phuoc}, \citenamefont {Malka},\ and\ \citenamefont
  {Thaury}}]{Smartsev2019}%
  \BibitemOpen
  \bibfield  {author} {\bibinfo {author} {\bibfnamefont {S.}~\bibnamefont
  {Smartsev}}, \bibinfo {author} {\bibfnamefont {C.}~\bibnamefont
  {Caizergues}}, \bibinfo {author} {\bibfnamefont {K.}~\bibnamefont
  {Oubrerie}}, \bibinfo {author} {\bibfnamefont {J.}~\bibnamefont {Gautier}},
  \bibinfo {author} {\bibfnamefont {J.-P.}\ \bibnamefont {Goddet}}, \bibinfo
  {author} {\bibfnamefont {A.}~\bibnamefont {Tafzi}}, \bibinfo {author}
  {\bibfnamefont {K.~T.}\ \bibnamefont {Phuoc}}, \bibinfo {author}
  {\bibfnamefont {V.}~\bibnamefont {Malka}}, \ and\ \bibinfo {author}
  {\bibfnamefont {C.}~\bibnamefont {Thaury}},\ }\href {\doibase
  10.1364/OL.44.003414} {\bibfield  {journal} {\bibinfo  {journal} {Optics
  Letters}\ }\textbf {\bibinfo {volume} {44}},\ \bibinfo {pages} {3414}
  (\bibinfo {year} {2019})}\BibitemShut {NoStop}%
\bibitem [{\citenamefont {Oubrerie}\ \emph {et~al.}(2022)\citenamefont
  {Oubrerie}, \citenamefont {Andriyash}, \citenamefont {Lahaye}, \citenamefont
  {Smartsev}, \citenamefont {Malka},\ and\ \citenamefont
  {Thaury}}]{Oubrerie2022}%
  \BibitemOpen
  \bibfield  {author} {\bibinfo {author} {\bibfnamefont {K.}~\bibnamefont
  {Oubrerie}}, \bibinfo {author} {\bibfnamefont {I.~A.}\ \bibnamefont
  {Andriyash}}, \bibinfo {author} {\bibfnamefont {R.}~\bibnamefont {Lahaye}},
  \bibinfo {author} {\bibfnamefont {S.}~\bibnamefont {Smartsev}}, \bibinfo
  {author} {\bibfnamefont {V.}~\bibnamefont {Malka}}, \ and\ \bibinfo {author}
  {\bibfnamefont {C.}~\bibnamefont {Thaury}},\ }\href {\doibase
  10.1088/2040-8986/ac57d2} {\bibfield  {journal} {\bibinfo  {journal} {Journal
  of Optics}\ }\textbf {\bibinfo {volume} {24}},\ \bibinfo {pages} {045503}
  (\bibinfo {year} {2022})}\BibitemShut {NoStop}%
\bibitem [{\citenamefont {Pigeon}\ \emph {et~al.}(2023)\citenamefont {Pigeon},
  \citenamefont {Franke}, \citenamefont {Chong}, \citenamefont {Katz},
  \citenamefont {Boni}, \citenamefont {Dorrer}, \citenamefont {Palastro},\ and\
  \citenamefont {Froula}}]{Pigeon2023}%
  \BibitemOpen
  \bibfield  {author} {\bibinfo {author} {\bibfnamefont {J.~J.}\ \bibnamefont
  {Pigeon}}, \bibinfo {author} {\bibfnamefont {P.}~\bibnamefont {Franke}},
  \bibinfo {author} {\bibfnamefont {M.~L.~P.}\ \bibnamefont {Chong}}, \bibinfo
  {author} {\bibfnamefont {J.}~\bibnamefont {Katz}}, \bibinfo {author}
  {\bibfnamefont {R.}~\bibnamefont {Boni}}, \bibinfo {author} {\bibfnamefont
  {C.}~\bibnamefont {Dorrer}}, \bibinfo {author} {\bibfnamefont {J.~P.}\
  \bibnamefont {Palastro}}, \ and\ \bibinfo {author} {\bibfnamefont {D.~H.}\
  \bibnamefont {Froula}}\ }(\bibinfo  {publisher} {Optica Publishing Group},\
  \bibinfo {year} {2023})\ p.\ \bibinfo {pages} {FW3M.1}\BibitemShut {NoStop}%
\bibitem [{\citenamefont {Caizergues}\ \emph {et~al.}(2020)\citenamefont
  {Caizergues}, \citenamefont {Smartsev}, \citenamefont {Malka},\ and\
  \citenamefont {Thaury}}]{Caizergues2020}%
  \BibitemOpen
  \bibfield  {author} {\bibinfo {author} {\bibfnamefont {C.}~\bibnamefont
  {Caizergues}}, \bibinfo {author} {\bibfnamefont {S.}~\bibnamefont
  {Smartsev}}, \bibinfo {author} {\bibfnamefont {V.}~\bibnamefont {Malka}}, \
  and\ \bibinfo {author} {\bibfnamefont {C.}~\bibnamefont {Thaury}},\ }\href
  {\doibase 10.1038/s41566-020-0657-2} {\bibfield  {journal} {\bibinfo
  {journal} {Nature Photonics}\ }\textbf {\bibinfo {volume} {14}},\ \bibinfo
  {pages} {475} (\bibinfo {year} {2020})}\BibitemShut {NoStop}%
\bibitem [{\citenamefont {Geng}\ \emph {et~al.}(2022)\citenamefont {Geng},
  \citenamefont {Chen}, \citenamefont {Zhu}, \citenamefont {Liu}, \citenamefont
  {Sheng},\ and\ \citenamefont {Zhang}}]{Geng2022}%
  \BibitemOpen
  \bibfield  {author} {\bibinfo {author} {\bibfnamefont {P.-F.}\ \bibnamefont
  {Geng}}, \bibinfo {author} {\bibfnamefont {M.}~\bibnamefont {Chen}}, \bibinfo
  {author} {\bibfnamefont {X.-Z.}\ \bibnamefont {Zhu}}, \bibinfo {author}
  {\bibfnamefont {W.-Y.}\ \bibnamefont {Liu}}, \bibinfo {author} {\bibfnamefont
  {Z.-M.}\ \bibnamefont {Sheng}}, \ and\ \bibinfo {author} {\bibfnamefont
  {J.}~\bibnamefont {Zhang}},\ }\href {\doibase 10.1063/5.0109643} {\bibfield
  {journal} {\bibinfo  {journal} {Physics of Plasmas}\ }\textbf {\bibinfo
  {volume} {29}},\ \bibinfo {pages} {112301} (\bibinfo {year}
  {2022})}\BibitemShut {NoStop}%
\bibitem [{\citenamefont {Geng}\ \emph {et~al.}(2023)\citenamefont {Geng},
  \citenamefont {Chen}, \citenamefont {An}, \citenamefont {Liu}, \citenamefont
  {Zhu}, \citenamefont {Li}, \citenamefont {Li},\ and\ \citenamefont
  {Sheng}}]{Geng2023}%
  \BibitemOpen
  \bibfield  {author} {\bibinfo {author} {\bibfnamefont {P.-F.}\ \bibnamefont
  {Geng}}, \bibinfo {author} {\bibfnamefont {M.}~\bibnamefont {Chen}}, \bibinfo
  {author} {\bibfnamefont {X.-Y.}\ \bibnamefont {An}}, \bibinfo {author}
  {\bibfnamefont {W.-Y.}\ \bibnamefont {Liu}}, \bibinfo {author} {\bibfnamefont
  {X.-Z.}\ \bibnamefont {Zhu}}, \bibinfo {author} {\bibfnamefont {J.-L.}\
  \bibnamefont {Li}}, \bibinfo {author} {\bibfnamefont {B.-Y.}\ \bibnamefont
  {Li}}, \ and\ \bibinfo {author} {\bibfnamefont {Z.-M.}\ \bibnamefont
  {Sheng}},\ }\href {\doibase 10.1088/1674-1056/acae79} {\bibfield  {journal}
  {\bibinfo  {journal} {Chinese Physics B}\ }\textbf {\bibinfo {volume} {32}},\
  \bibinfo {pages} {044101} (\bibinfo {year} {2023})}\BibitemShut {NoStop}%
\bibitem [{\citenamefont {Palastro}\ \emph {et~al.}(2021)\citenamefont
  {Palastro}, \citenamefont {Malaca}, \citenamefont {Vieira}, \citenamefont
  {Ramsey}, \citenamefont {Simpson}, \citenamefont {Franke}, \citenamefont
  {Shaw},\ and\ \citenamefont {Froula}}]{Palastro2021}%
  \BibitemOpen
  \bibfield  {author} {\bibinfo {author} {\bibfnamefont {J.~P.}\ \bibnamefont
  {Palastro}}, \bibinfo {author} {\bibfnamefont {B.}~\bibnamefont {Malaca}},
  \bibinfo {author} {\bibfnamefont {J.}~\bibnamefont {Vieira}}, \bibinfo
  {author} {\bibfnamefont {D.}~\bibnamefont {Ramsey}}, \bibinfo {author}
  {\bibfnamefont {T.~T.}\ \bibnamefont {Simpson}}, \bibinfo {author}
  {\bibfnamefont {P.}~\bibnamefont {Franke}}, \bibinfo {author} {\bibfnamefont
  {J.~L.}\ \bibnamefont {Shaw}}, \ and\ \bibinfo {author} {\bibfnamefont
  {D.~H.}\ \bibnamefont {Froula}},\ }\href {\doibase 10.1063/5.0036627}
  {\bibfield  {journal} {\bibinfo  {journal} {Physics of Plasmas}\ }\textbf
  {\bibinfo {volume} {28}},\ \bibinfo {pages} {013109} (\bibinfo {year}
  {2021})}\BibitemShut {NoStop}%
\bibitem [{\citenamefont {Froula}\ \emph {et~al.}(2009)\citenamefont {Froula},
  \citenamefont {Clayton}, \citenamefont {Döppner}, \citenamefont {Marsh},
  \citenamefont {Barty}, \citenamefont {Divol}, \citenamefont {Fonseca},
  \citenamefont {Glenzer}, \citenamefont {Joshi}, \citenamefont {Lu},
  \citenamefont {Martins}, \citenamefont {Michel}, \citenamefont {Mori},
  \citenamefont {Palastro}, \citenamefont {Pollock}, \citenamefont {Pak},
  \citenamefont {Ralph}, \citenamefont {Ross}, \citenamefont {Siders},
  \citenamefont {Silva},\ and\ \citenamefont {Wang}}]{Froula2009}%
  \BibitemOpen
  \bibfield  {author} {\bibinfo {author} {\bibfnamefont {D.~H.}\ \bibnamefont
  {Froula}}, \bibinfo {author} {\bibfnamefont {C.~E.}\ \bibnamefont {Clayton}},
  \bibinfo {author} {\bibfnamefont {T.}~\bibnamefont {Döppner}}, \bibinfo
  {author} {\bibfnamefont {K.~A.}\ \bibnamefont {Marsh}}, \bibinfo {author}
  {\bibfnamefont {C.~P.~J.}\ \bibnamefont {Barty}}, \bibinfo {author}
  {\bibfnamefont {L.}~\bibnamefont {Divol}}, \bibinfo {author} {\bibfnamefont
  {R.~A.}\ \bibnamefont {Fonseca}}, \bibinfo {author} {\bibfnamefont {S.~H.}\
  \bibnamefont {Glenzer}}, \bibinfo {author} {\bibfnamefont {C.}~\bibnamefont
  {Joshi}}, \bibinfo {author} {\bibfnamefont {W.}~\bibnamefont {Lu}}, \bibinfo
  {author} {\bibfnamefont {S.~F.}\ \bibnamefont {Martins}}, \bibinfo {author}
  {\bibfnamefont {P.}~\bibnamefont {Michel}}, \bibinfo {author} {\bibfnamefont
  {W.~B.}\ \bibnamefont {Mori}}, \bibinfo {author} {\bibfnamefont {J.~P.}\
  \bibnamefont {Palastro}}, \bibinfo {author} {\bibfnamefont {B.~B.}\
  \bibnamefont {Pollock}}, \bibinfo {author} {\bibfnamefont {A.}~\bibnamefont
  {Pak}}, \bibinfo {author} {\bibfnamefont {J.~E.}\ \bibnamefont {Ralph}},
  \bibinfo {author} {\bibfnamefont {J.~S.}\ \bibnamefont {Ross}}, \bibinfo
  {author} {\bibfnamefont {C.~W.}\ \bibnamefont {Siders}}, \bibinfo {author}
  {\bibfnamefont {L.~O.}\ \bibnamefont {Silva}}, \ and\ \bibinfo {author}
  {\bibfnamefont {T.}~\bibnamefont {Wang}},\ }\href {\doibase
  10.1103/PhysRevLett.103.215006} {\bibfield  {journal} {\bibinfo  {journal}
  {Physical Review Letters}\ }\textbf {\bibinfo {volume} {103}},\ \bibinfo
  {pages} {215006} (\bibinfo {year} {2009})}\BibitemShut {NoStop}%
\bibitem [{\citenamefont {Mangles}\ \emph {et~al.}(2012)\citenamefont
  {Mangles}, \citenamefont {Genoud}, \citenamefont {Bloom}, \citenamefont
  {Burza}, \citenamefont {Najmudin}, \citenamefont {Persson}, \citenamefont
  {Svensson}, \citenamefont {Thomas},\ and\ \citenamefont
  {Wahlström}}]{Mangles2012}%
  \BibitemOpen
  \bibfield  {author} {\bibinfo {author} {\bibfnamefont {S.~P.~D.}\
  \bibnamefont {Mangles}}, \bibinfo {author} {\bibfnamefont {G.}~\bibnamefont
  {Genoud}}, \bibinfo {author} {\bibfnamefont {M.~S.}\ \bibnamefont {Bloom}},
  \bibinfo {author} {\bibfnamefont {M.}~\bibnamefont {Burza}}, \bibinfo
  {author} {\bibfnamefont {Z.}~\bibnamefont {Najmudin}}, \bibinfo {author}
  {\bibfnamefont {A.}~\bibnamefont {Persson}}, \bibinfo {author} {\bibfnamefont
  {K.}~\bibnamefont {Svensson}}, \bibinfo {author} {\bibfnamefont {A.~G.~R.}\
  \bibnamefont {Thomas}}, \ and\ \bibinfo {author} {\bibfnamefont {C.-G.}\
  \bibnamefont {Wahlström}},\ }\href {\doibase 10.1103/PhysRevSTAB.15.011302}
  {\bibfield  {journal} {\bibinfo  {journal} {Physical Review Special Topics -
  Accelerators and Beams}\ }\textbf {\bibinfo {volume} {15}},\ \bibinfo {pages}
  {011302} (\bibinfo {year} {2012})}\BibitemShut {NoStop}%
\bibitem [{\citenamefont {Chen}\ \emph {et~al.}(2006)\citenamefont {Chen},
  \citenamefont {Sheng}, \citenamefont {Ma},\ and\ \citenamefont
  {Zhang}}]{Chen2006}%
  \BibitemOpen
  \bibfield  {author} {\bibinfo {author} {\bibfnamefont {M.}~\bibnamefont
  {Chen}}, \bibinfo {author} {\bibfnamefont {Z.-M.}\ \bibnamefont {Sheng}},
  \bibinfo {author} {\bibfnamefont {Y.-Y.}\ \bibnamefont {Ma}}, \ and\ \bibinfo
  {author} {\bibfnamefont {J.}~\bibnamefont {Zhang}},\ }\href {\doibase
  10.1063/1.2179194} {\bibfield  {journal} {\bibinfo  {journal} {Journal of
  Applied Physics}\ }\textbf {\bibinfo {volume} {99}} (\bibinfo {year}
  {2006}),\ 10.1063/1.2179194}\BibitemShut {NoStop}%
\bibitem [{\citenamefont {Pak}\ \emph {et~al.}(2010)\citenamefont {Pak},
  \citenamefont {Marsh}, \citenamefont {Martins}, \citenamefont {Lu},
  \citenamefont {Mori},\ and\ \citenamefont {Joshi}}]{Pak2010}%
  \BibitemOpen
  \bibfield  {author} {\bibinfo {author} {\bibfnamefont {A.}~\bibnamefont
  {Pak}}, \bibinfo {author} {\bibfnamefont {K.~A.}\ \bibnamefont {Marsh}},
  \bibinfo {author} {\bibfnamefont {S.~F.}\ \bibnamefont {Martins}}, \bibinfo
  {author} {\bibfnamefont {W.}~\bibnamefont {Lu}}, \bibinfo {author}
  {\bibfnamefont {W.~B.}\ \bibnamefont {Mori}}, \ and\ \bibinfo {author}
  {\bibfnamefont {C.}~\bibnamefont {Joshi}},\ }\href {\doibase
  10.1103/PhysRevLett.104.025003} {\bibfield  {journal} {\bibinfo  {journal}
  {Physical Review Letters}\ }\textbf {\bibinfo {volume} {104}},\ \bibinfo
  {pages} {025003} (\bibinfo {year} {2010})}\BibitemShut {NoStop}%
\bibitem [{\citenamefont {McGuffey}\ \emph {et~al.}(2010)\citenamefont
  {McGuffey}, \citenamefont {Thomas}, \citenamefont {Schumaker}, \citenamefont
  {Matsuoka}, \citenamefont {Chvykov}, \citenamefont {Dollar}, \citenamefont
  {Kalintchenko}, \citenamefont {Yanovsky}, \citenamefont {Maksimchuk},
  \citenamefont {Krushelnick}, \citenamefont {Bychenkov}, \citenamefont
  {Glazyrin},\ and\ \citenamefont {Karpeev}}]{McGuffey2010}%
  \BibitemOpen
  \bibfield  {author} {\bibinfo {author} {\bibfnamefont {C.}~\bibnamefont
  {McGuffey}}, \bibinfo {author} {\bibfnamefont {A.~G.~R.}\ \bibnamefont
  {Thomas}}, \bibinfo {author} {\bibfnamefont {W.}~\bibnamefont {Schumaker}},
  \bibinfo {author} {\bibfnamefont {T.}~\bibnamefont {Matsuoka}}, \bibinfo
  {author} {\bibfnamefont {V.}~\bibnamefont {Chvykov}}, \bibinfo {author}
  {\bibfnamefont {F.~J.}\ \bibnamefont {Dollar}}, \bibinfo {author}
  {\bibfnamefont {G.}~\bibnamefont {Kalintchenko}}, \bibinfo {author}
  {\bibfnamefont {V.}~\bibnamefont {Yanovsky}}, \bibinfo {author}
  {\bibfnamefont {A.}~\bibnamefont {Maksimchuk}}, \bibinfo {author}
  {\bibfnamefont {K.}~\bibnamefont {Krushelnick}}, \bibinfo {author}
  {\bibfnamefont {V.~Y.}\ \bibnamefont {Bychenkov}}, \bibinfo {author}
  {\bibfnamefont {I.~V.}\ \bibnamefont {Glazyrin}}, \ and\ \bibinfo {author}
  {\bibfnamefont {A.~V.}\ \bibnamefont {Karpeev}},\ }\href {\doibase
  10.1103/PhysRevLett.104.025004} {\bibfield  {journal} {\bibinfo  {journal}
  {Physical Review Letters}\ }\textbf {\bibinfo {volume} {104}},\ \bibinfo
  {pages} {025004} (\bibinfo {year} {2010})}\BibitemShut {NoStop}%
\bibitem [{\citenamefont {Pollock}\ \emph {et~al.}(2011)\citenamefont
  {Pollock}, \citenamefont {Clayton}, \citenamefont {Ralph}, \citenamefont
  {Albert}, \citenamefont {Davidson}, \citenamefont {Divol}, \citenamefont
  {Filip}, \citenamefont {Glenzer}, \citenamefont {Herpoldt}, \citenamefont
  {Lu}, \citenamefont {Marsh}, \citenamefont {Meinecke}, \citenamefont {Mori},
  \citenamefont {Pak}, \citenamefont {Rensink}, \citenamefont {Ross},
  \citenamefont {Shaw}, \citenamefont {Tynan}, \citenamefont {Joshi},\ and\
  \citenamefont {Froula}}]{Pollock2011}%
  \BibitemOpen
  \bibfield  {author} {\bibinfo {author} {\bibfnamefont {B.~B.}\ \bibnamefont
  {Pollock}}, \bibinfo {author} {\bibfnamefont {C.~E.}\ \bibnamefont
  {Clayton}}, \bibinfo {author} {\bibfnamefont {J.~E.}\ \bibnamefont {Ralph}},
  \bibinfo {author} {\bibfnamefont {F.}~\bibnamefont {Albert}}, \bibinfo
  {author} {\bibfnamefont {A.}~\bibnamefont {Davidson}}, \bibinfo {author}
  {\bibfnamefont {L.}~\bibnamefont {Divol}}, \bibinfo {author} {\bibfnamefont
  {C.}~\bibnamefont {Filip}}, \bibinfo {author} {\bibfnamefont {S.~H.}\
  \bibnamefont {Glenzer}}, \bibinfo {author} {\bibfnamefont {K.}~\bibnamefont
  {Herpoldt}}, \bibinfo {author} {\bibfnamefont {W.}~\bibnamefont {Lu}},
  \bibinfo {author} {\bibfnamefont {K.~A.}\ \bibnamefont {Marsh}}, \bibinfo
  {author} {\bibfnamefont {J.}~\bibnamefont {Meinecke}}, \bibinfo {author}
  {\bibfnamefont {W.~B.}\ \bibnamefont {Mori}}, \bibinfo {author}
  {\bibfnamefont {A.}~\bibnamefont {Pak}}, \bibinfo {author} {\bibfnamefont
  {T.~C.}\ \bibnamefont {Rensink}}, \bibinfo {author} {\bibfnamefont {J.~S.}\
  \bibnamefont {Ross}}, \bibinfo {author} {\bibfnamefont {J.}~\bibnamefont
  {Shaw}}, \bibinfo {author} {\bibfnamefont {G.~R.}\ \bibnamefont {Tynan}},
  \bibinfo {author} {\bibfnamefont {C.}~\bibnamefont {Joshi}}, \ and\ \bibinfo
  {author} {\bibfnamefont {D.~H.}\ \bibnamefont {Froula}},\ }\href {\doibase
  10.1103/PhysRevLett.107.045001} {\bibfield  {journal} {\bibinfo  {journal}
  {Physical Review Letters}\ }\textbf {\bibinfo {volume} {107}},\ \bibinfo
  {pages} {045001} (\bibinfo {year} {2011})}\BibitemShut {NoStop}%
\bibitem [{\citenamefont {Xu}\ \emph {et~al.}(2014)\citenamefont {Xu},
  \citenamefont {Wu}, \citenamefont {Zhang}, \citenamefont {Li}, \citenamefont
  {Wan}, \citenamefont {Hua}, \citenamefont {Pai}, \citenamefont {Lu},
  \citenamefont {Yu}, \citenamefont {Joshi},\ and\ \citenamefont
  {Mori}}]{Xu2014}%
  \BibitemOpen
  \bibfield  {author} {\bibinfo {author} {\bibfnamefont {X.}~\bibnamefont
  {Xu}}, \bibinfo {author} {\bibfnamefont {Y.}~\bibnamefont {Wu}}, \bibinfo
  {author} {\bibfnamefont {C.}~\bibnamefont {Zhang}}, \bibinfo {author}
  {\bibfnamefont {F.}~\bibnamefont {Li}}, \bibinfo {author} {\bibfnamefont
  {Y.}~\bibnamefont {Wan}}, \bibinfo {author} {\bibfnamefont {J.}~\bibnamefont
  {Hua}}, \bibinfo {author} {\bibfnamefont {C.-H.}\ \bibnamefont {Pai}},
  \bibinfo {author} {\bibfnamefont {W.}~\bibnamefont {Lu}}, \bibinfo {author}
  {\bibfnamefont {P.}~\bibnamefont {Yu}}, \bibinfo {author} {\bibfnamefont
  {C.}~\bibnamefont {Joshi}}, \ and\ \bibinfo {author} {\bibfnamefont
  {W.}~\bibnamefont {Mori}},\ }\href {\doibase 10.1103/PhysRevSTAB.17.061301}
  {\bibfield  {journal} {\bibinfo  {journal} {Physical Review Special Topics -
  Accelerators and Beams}\ }\textbf {\bibinfo {volume} {17}},\ \bibinfo {pages}
  {061301} (\bibinfo {year} {2014})}\BibitemShut {NoStop}%
\bibitem [{\citenamefont {Zhang}\ \emph {et~al.}(2016)\citenamefont {Zhang},
  \citenamefont {Khudik}, \citenamefont {Pukhov},\ and\ \citenamefont
  {Shvets}}]{Zhang2016}%
  \BibitemOpen
  \bibfield  {author} {\bibinfo {author} {\bibfnamefont {X.}~\bibnamefont
  {Zhang}}, \bibinfo {author} {\bibfnamefont {V.~N.}\ \bibnamefont {Khudik}},
  \bibinfo {author} {\bibfnamefont {A.}~\bibnamefont {Pukhov}}, \ and\ \bibinfo
  {author} {\bibfnamefont {G.}~\bibnamefont {Shvets}},\ }\href {\doibase
  10.1088/0741-3335/58/3/034011} {\bibfield  {journal} {\bibinfo  {journal}
  {Plasma Physics and Controlled Fusion}\ }\textbf {\bibinfo {volume} {58}},\
  \bibinfo {pages} {034011} (\bibinfo {year} {2016})}\BibitemShut {NoStop}%
\bibitem [{\citenamefont {Oz}\ \emph {et~al.}(2007)\citenamefont {Oz},
  \citenamefont {Deng}, \citenamefont {Katsouleas}, \citenamefont {Muggli},
  \citenamefont {Barnes}, \citenamefont {Blumenfeld}, \citenamefont {Decker},
  \citenamefont {Emma}, \citenamefont {Hogan}, \citenamefont {Ischebeck},
  \citenamefont {Iverson}, \citenamefont {Kirby}, \citenamefont {Krejcik},
  \citenamefont {O’Connell}, \citenamefont {Siemann}, \citenamefont {Walz},
  \citenamefont {Auerbach}, \citenamefont {Clayton}, \citenamefont {Huang},
  \citenamefont {Johnson}, \citenamefont {Joshi}, \citenamefont {Lu},
  \citenamefont {Marsh}, \citenamefont {Mori},\ and\ \citenamefont
  {Zhou}}]{Oz2007}%
  \BibitemOpen
  \bibfield  {author} {\bibinfo {author} {\bibfnamefont {E.}~\bibnamefont
  {Oz}}, \bibinfo {author} {\bibfnamefont {S.}~\bibnamefont {Deng}}, \bibinfo
  {author} {\bibfnamefont {T.}~\bibnamefont {Katsouleas}}, \bibinfo {author}
  {\bibfnamefont {P.}~\bibnamefont {Muggli}}, \bibinfo {author} {\bibfnamefont
  {C.~D.}\ \bibnamefont {Barnes}}, \bibinfo {author} {\bibfnamefont
  {I.}~\bibnamefont {Blumenfeld}}, \bibinfo {author} {\bibfnamefont {F.~J.}\
  \bibnamefont {Decker}}, \bibinfo {author} {\bibfnamefont {P.}~\bibnamefont
  {Emma}}, \bibinfo {author} {\bibfnamefont {M.~J.}\ \bibnamefont {Hogan}},
  \bibinfo {author} {\bibfnamefont {R.}~\bibnamefont {Ischebeck}}, \bibinfo
  {author} {\bibfnamefont {R.~H.}\ \bibnamefont {Iverson}}, \bibinfo {author}
  {\bibfnamefont {N.}~\bibnamefont {Kirby}}, \bibinfo {author} {\bibfnamefont
  {P.}~\bibnamefont {Krejcik}}, \bibinfo {author} {\bibfnamefont
  {C.}~\bibnamefont {O’Connell}}, \bibinfo {author} {\bibfnamefont {R.~H.}\
  \bibnamefont {Siemann}}, \bibinfo {author} {\bibfnamefont {D.}~\bibnamefont
  {Walz}}, \bibinfo {author} {\bibfnamefont {D.}~\bibnamefont {Auerbach}},
  \bibinfo {author} {\bibfnamefont {C.~E.}\ \bibnamefont {Clayton}}, \bibinfo
  {author} {\bibfnamefont {C.}~\bibnamefont {Huang}}, \bibinfo {author}
  {\bibfnamefont {D.~K.}\ \bibnamefont {Johnson}}, \bibinfo {author}
  {\bibfnamefont {C.}~\bibnamefont {Joshi}}, \bibinfo {author} {\bibfnamefont
  {W.}~\bibnamefont {Lu}}, \bibinfo {author} {\bibfnamefont {K.~A.}\
  \bibnamefont {Marsh}}, \bibinfo {author} {\bibfnamefont {W.~B.}\ \bibnamefont
  {Mori}}, \ and\ \bibinfo {author} {\bibfnamefont {M.}~\bibnamefont {Zhou}},\
  }\href {\doibase 10.1103/PhysRevLett.98.084801} {\bibfield  {journal}
  {\bibinfo  {journal} {Physical Review Letters}\ }\textbf {\bibinfo {volume}
  {98}},\ \bibinfo {pages} {084801} (\bibinfo {year} {2007})}\BibitemShut
  {NoStop}%
\bibitem [{\citenamefont {Sun}\ \emph {et~al.}(1987)\citenamefont {Sun},
  \citenamefont {Ott}, \citenamefont {Lee},\ and\ \citenamefont
  {Guzdar}}]{Sun1987}%
  \BibitemOpen
  \bibfield  {author} {\bibinfo {author} {\bibfnamefont {G.-Z.}\ \bibnamefont
  {Sun}}, \bibinfo {author} {\bibfnamefont {E.}~\bibnamefont {Ott}}, \bibinfo
  {author} {\bibfnamefont {Y.~C.}\ \bibnamefont {Lee}}, \ and\ \bibinfo
  {author} {\bibfnamefont {P.}~\bibnamefont {Guzdar}},\ }\href {\doibase
  10.1063/1.866349} {\bibfield  {journal} {\bibinfo  {journal} {Physics of
  Fluids}\ }\textbf {\bibinfo {volume} {30}},\ \bibinfo {pages} {526} (\bibinfo
  {year} {1987})}\BibitemShut {NoStop}%
\bibitem [{\citenamefont {Katsouleas}\ \emph {et~al.}(1987)\citenamefont
  {Katsouleas}, \citenamefont {Wilks}, \citenamefont {Chen}, \citenamefont
  {Dawson},\ and\ \citenamefont {Su}}]{Katsouleas1987}%
  \BibitemOpen
  \bibfield  {author} {\bibinfo {author} {\bibfnamefont {T.}~\bibnamefont
  {Katsouleas}}, \bibinfo {author} {\bibfnamefont {S.}~\bibnamefont {Wilks}},
  \bibinfo {author} {\bibfnamefont {P.}~\bibnamefont {Chen}}, \bibinfo {author}
  {\bibfnamefont {J.~M.}\ \bibnamefont {Dawson}}, \ and\ \bibinfo {author}
  {\bibfnamefont {J.~J.}\ \bibnamefont {Su}},\ }\href
  {http://cds.cern.ch/record/898463/files/p81. pdf} {\bibfield  {journal}
  {\bibinfo  {journal} {Particle Accelerators}\ }\textbf {\bibinfo {volume}
  {22}},\ \bibinfo {pages} {81} (\bibinfo {year} {1987})}\BibitemShut {NoStop}%
\bibitem [{\citenamefont {Tzoufras}\ \emph {et~al.}(2008)\citenamefont
  {Tzoufras}, \citenamefont {Lu}, \citenamefont {Tsung}, \citenamefont {Huang},
  \citenamefont {Mori}, \citenamefont {Katsouleas}, \citenamefont {Vieira},
  \citenamefont {Fonseca},\ and\ \citenamefont {Silva}}]{Tzoufras2008}%
  \BibitemOpen
  \bibfield  {author} {\bibinfo {author} {\bibfnamefont {M.}~\bibnamefont
  {Tzoufras}}, \bibinfo {author} {\bibfnamefont {W.}~\bibnamefont {Lu}},
  \bibinfo {author} {\bibfnamefont {F.~S.}\ \bibnamefont {Tsung}}, \bibinfo
  {author} {\bibfnamefont {C.}~\bibnamefont {Huang}}, \bibinfo {author}
  {\bibfnamefont {W.~B.}\ \bibnamefont {Mori}}, \bibinfo {author}
  {\bibfnamefont {T.}~\bibnamefont {Katsouleas}}, \bibinfo {author}
  {\bibfnamefont {J.}~\bibnamefont {Vieira}}, \bibinfo {author} {\bibfnamefont
  {R.~A.}\ \bibnamefont {Fonseca}}, \ and\ \bibinfo {author} {\bibfnamefont
  {L.~O.}\ \bibnamefont {Silva}},\ }\href {\doibase
  10.1103/PhysRevLett.101.145002} {\bibfield  {journal} {\bibinfo  {journal}
  {Physical Review Letters}\ }\textbf {\bibinfo {volume} {101}},\ \bibinfo
  {pages} {145002} (\bibinfo {year} {2008})}\BibitemShut {NoStop}%
\bibitem [{\citenamefont {Dalichaouch}\ \emph {et~al.}(2021)\citenamefont
  {Dalichaouch}, \citenamefont {Xu}, \citenamefont {Tableman}, \citenamefont
  {Li}, \citenamefont {Tsung},\ and\ \citenamefont {Mori}}]{Dalichaouch2021}%
  \BibitemOpen
  \bibfield  {author} {\bibinfo {author} {\bibfnamefont {T.~N.}\ \bibnamefont
  {Dalichaouch}}, \bibinfo {author} {\bibfnamefont {X.~L.}\ \bibnamefont {Xu}},
  \bibinfo {author} {\bibfnamefont {A.}~\bibnamefont {Tableman}}, \bibinfo
  {author} {\bibfnamefont {F.}~\bibnamefont {Li}}, \bibinfo {author}
  {\bibfnamefont {F.~S.}\ \bibnamefont {Tsung}}, \ and\ \bibinfo {author}
  {\bibfnamefont {W.~B.}\ \bibnamefont {Mori}},\ }\href {\doibase
  10.1063/5.0051282} {\bibfield  {journal} {\bibinfo  {journal} {Physics of
  Plasmas}\ }\textbf {\bibinfo {volume} {28}},\ \bibinfo {pages} {063103}
  (\bibinfo {year} {2021})}\BibitemShut {NoStop}%
\bibitem [{\citenamefont {Sun}\ \emph {et~al.}(2015)\citenamefont {Sun},
  \citenamefont {Salter},\ and\ \citenamefont {Booth}}]{Sun2015}%
  \BibitemOpen
  \bibfield  {author} {\bibinfo {author} {\bibfnamefont {B.}~\bibnamefont
  {Sun}}, \bibinfo {author} {\bibfnamefont {P.~S.}\ \bibnamefont {Salter}}, \
  and\ \bibinfo {author} {\bibfnamefont {M.~J.}\ \bibnamefont {Booth}},\ }\href
  {\doibase 10.1364/OE.23.019348} {\bibfield  {journal} {\bibinfo  {journal}
  {Optics Express}\ }\textbf {\bibinfo {volume} {23}},\ \bibinfo {pages}
  {19348} (\bibinfo {year} {2015})}\BibitemShut {NoStop}%
\bibitem [{\citenamefont {Li}\ and\ \citenamefont
  {Kawanaka}(2020{\natexlab{a}})}]{Li2020}%
  \BibitemOpen
  \bibfield  {author} {\bibinfo {author} {\bibfnamefont {Z.}~\bibnamefont
  {Li}}\ and\ \bibinfo {author} {\bibfnamefont {J.}~\bibnamefont {Kawanaka}},\
  }\href {\doibase 10.1038/s42005-020-00481-4} {\bibfield  {journal} {\bibinfo
  {journal} {Communications Physics}\ }\textbf {\bibinfo {volume} {3}},\
  \bibinfo {pages} {211} (\bibinfo {year} {2020}{\natexlab{a}})}\BibitemShut
  {NoStop}%
\bibitem [{\citenamefont {Tzoufras}\ \emph {et~al.}(2012)\citenamefont
  {Tzoufras}, \citenamefont {Huang}, \citenamefont {Cooley}, \citenamefont
  {Tsung}, \citenamefont {Vieira},\ and\ \citenamefont {Mori}}]{Tzoufras2012}%
  \BibitemOpen
  \bibfield  {author} {\bibinfo {author} {\bibfnamefont {M.}~\bibnamefont
  {Tzoufras}}, \bibinfo {author} {\bibfnamefont {C.}~\bibnamefont {Huang}},
  \bibinfo {author} {\bibfnamefont {J.~H.}\ \bibnamefont {Cooley}}, \bibinfo
  {author} {\bibfnamefont {F.~S.}\ \bibnamefont {Tsung}}, \bibinfo {author}
  {\bibfnamefont {J.}~\bibnamefont {Vieira}}, \ and\ \bibinfo {author}
  {\bibfnamefont {W.~B.}\ \bibnamefont {Mori}},\ }\href {\doibase
  10.1017/S0022377812000232} {\bibfield  {journal} {\bibinfo  {journal}
  {Journal of Plasma Physics}\ }\textbf {\bibinfo {volume} {78}},\ \bibinfo
  {pages} {401} (\bibinfo {year} {2012})}\BibitemShut {NoStop}%
\bibitem [{\citenamefont {Pierce}\ \emph {et~al.}(2023)\citenamefont {Pierce},
  \citenamefont {Palastro}, \citenamefont {Li}, \citenamefont {Malaca},
  \citenamefont {Ramsey}, \citenamefont {Vieira}, \citenamefont {Weichman},\
  and\ \citenamefont {Mori}}]{Pierce2023}%
  \BibitemOpen
  \bibfield  {author} {\bibinfo {author} {\bibfnamefont {J.~R.}\ \bibnamefont
  {Pierce}}, \bibinfo {author} {\bibfnamefont {J.~P.}\ \bibnamefont
  {Palastro}}, \bibinfo {author} {\bibfnamefont {F.}~\bibnamefont {Li}},
  \bibinfo {author} {\bibfnamefont {B.}~\bibnamefont {Malaca}}, \bibinfo
  {author} {\bibfnamefont {D.}~\bibnamefont {Ramsey}}, \bibinfo {author}
  {\bibfnamefont {J.}~\bibnamefont {Vieira}}, \bibinfo {author} {\bibfnamefont
  {K.}~\bibnamefont {Weichman}}, \ and\ \bibinfo {author} {\bibfnamefont
  {W.~B.}\ \bibnamefont {Mori}},\ }\href {\doibase
  10.1103/PhysRevResearch.5.013085} {\bibfield  {journal} {\bibinfo  {journal}
  {Physical Review Research}\ }\textbf {\bibinfo {volume} {5}},\ \bibinfo
  {pages} {013085} (\bibinfo {year} {2023})}\BibitemShut {NoStop}%
\bibitem [{\citenamefont {Yessenov}\ \emph {et~al.}(2022)\citenamefont
  {Yessenov}, \citenamefont {Free}, \citenamefont {Chen}, \citenamefont
  {Johnson}, \citenamefont {Lavery}, \citenamefont {Alonso},\ and\
  \citenamefont {Abouraddy}}]{Yessenov2022}%
  \BibitemOpen
  \bibfield  {author} {\bibinfo {author} {\bibfnamefont {M.}~\bibnamefont
  {Yessenov}}, \bibinfo {author} {\bibfnamefont {J.}~\bibnamefont {Free}},
  \bibinfo {author} {\bibfnamefont {Z.}~\bibnamefont {Chen}}, \bibinfo {author}
  {\bibfnamefont {E.~G.}\ \bibnamefont {Johnson}}, \bibinfo {author}
  {\bibfnamefont {M.~P.~J.}\ \bibnamefont {Lavery}}, \bibinfo {author}
  {\bibfnamefont {M.~A.}\ \bibnamefont {Alonso}}, \ and\ \bibinfo {author}
  {\bibfnamefont {A.~F.}\ \bibnamefont {Abouraddy}},\ }\href {\doibase
  10.1038/s41467-022-32240-0} {\bibfield  {journal} {\bibinfo  {journal}
  {Nature Communications}\ }\textbf {\bibinfo {volume} {13}},\ \bibinfo {pages}
  {4573} (\bibinfo {year} {2022})}\BibitemShut {NoStop}%
\bibitem [{\citenamefont {Simpson}\ \emph {et~al.}(2022)\citenamefont
  {Simpson}, \citenamefont {Ramsey}, \citenamefont {Franke}, \citenamefont
  {Weichman}, \citenamefont {Ambat}, \citenamefont {Turnbull}, \citenamefont
  {Froula},\ and\ \citenamefont {Palastro}}]{Simpson2022}%
  \BibitemOpen
  \bibfield  {author} {\bibinfo {author} {\bibfnamefont {T.~T.}\ \bibnamefont
  {Simpson}}, \bibinfo {author} {\bibfnamefont {D.}~\bibnamefont {Ramsey}},
  \bibinfo {author} {\bibfnamefont {P.}~\bibnamefont {Franke}}, \bibinfo
  {author} {\bibfnamefont {K.}~\bibnamefont {Weichman}}, \bibinfo {author}
  {\bibfnamefont {M.~V.}\ \bibnamefont {Ambat}}, \bibinfo {author}
  {\bibfnamefont {D.}~\bibnamefont {Turnbull}}, \bibinfo {author}
  {\bibfnamefont {D.~H.}\ \bibnamefont {Froula}}, \ and\ \bibinfo {author}
  {\bibfnamefont {J.~P.}\ \bibnamefont {Palastro}},\ }\href {\doibase
  10.1364/OE.451123} {\bibfield  {journal} {\bibinfo  {journal} {Optics
  Express}\ }\textbf {\bibinfo {volume} {30}},\ \bibinfo {pages} {9878}
  (\bibinfo {year} {2022})}\BibitemShut {NoStop}%
\bibitem [{\citenamefont {Li}\ and\ \citenamefont
  {Kawanaka}(2020{\natexlab{b}})}]{Li2020a}%
  \BibitemOpen
  \bibfield  {author} {\bibinfo {author} {\bibfnamefont {Z.}~\bibnamefont
  {Li}}\ and\ \bibinfo {author} {\bibfnamefont {J.}~\bibnamefont {Kawanaka}},\
  }\href {\doibase 10.1038/s41598-020-68478-1} {\bibfield  {journal} {\bibinfo
  {journal} {Scientific Reports}\ }\textbf {\bibinfo {volume} {10}},\ \bibinfo
  {pages} {11481} (\bibinfo {year} {2020}{\natexlab{b}})}\BibitemShut {NoStop}%
\bibitem [{\citenamefont {Lu}\ \emph {et~al.}(2006)\citenamefont {Lu},
  \citenamefont {Huang}, \citenamefont {Zhou}, \citenamefont {Mori},\ and\
  \citenamefont {Katsouleas}}]{Lu2006}%
  \BibitemOpen
  \bibfield  {author} {\bibinfo {author} {\bibfnamefont {W.}~\bibnamefont
  {Lu}}, \bibinfo {author} {\bibfnamefont {C.}~\bibnamefont {Huang}}, \bibinfo
  {author} {\bibfnamefont {M.}~\bibnamefont {Zhou}}, \bibinfo {author}
  {\bibfnamefont {W.~B.}\ \bibnamefont {Mori}}, \ and\ \bibinfo {author}
  {\bibfnamefont {T.}~\bibnamefont {Katsouleas}},\ }\href {\doibase
  10.1103/PhysRevLett.96.165002} {\bibfield  {journal} {\bibinfo  {journal}
  {Physical Review Letters}\ }\textbf {\bibinfo {volume} {96}},\ \bibinfo
  {pages} {165002} (\bibinfo {year} {2006})}\BibitemShut {NoStop}%
\bibitem [{\citenamefont {Palastro}\ \emph {et~al.}(2018)\citenamefont
  {Palastro}, \citenamefont {Turnbull}, \citenamefont {Bahk}, \citenamefont
  {Follett}, \citenamefont {Shaw}, \citenamefont {Haberberger}, \citenamefont
  {Bromage},\ and\ \citenamefont {Froula}}]{Palastro2018}%
  \BibitemOpen
  \bibfield  {author} {\bibinfo {author} {\bibfnamefont {J.~P.}\ \bibnamefont
  {Palastro}}, \bibinfo {author} {\bibfnamefont {D.}~\bibnamefont {Turnbull}},
  \bibinfo {author} {\bibfnamefont {S.-W.}\ \bibnamefont {Bahk}}, \bibinfo
  {author} {\bibfnamefont {R.~K.}\ \bibnamefont {Follett}}, \bibinfo {author}
  {\bibfnamefont {J.~L.}\ \bibnamefont {Shaw}}, \bibinfo {author}
  {\bibfnamefont {D.}~\bibnamefont {Haberberger}}, \bibinfo {author}
  {\bibfnamefont {J.}~\bibnamefont {Bromage}}, \ and\ \bibinfo {author}
  {\bibfnamefont {D.~H.}\ \bibnamefont {Froula}},\ }\href {\doibase
  10.1103/PhysRevA.97.033835} {\bibfield  {journal} {\bibinfo  {journal}
  {Physical Review A}\ }\textbf {\bibinfo {volume} {97}},\ \bibinfo {pages}
  {033835} (\bibinfo {year} {2018})}\BibitemShut {NoStop}%
\bibitem [{\citenamefont {Kolesik}\ \emph {et~al.}(2002)\citenamefont
  {Kolesik}, \citenamefont {Moloney},\ and\ \citenamefont
  {Mlejnek}}]{Kolesik2002}%
  \BibitemOpen
  \bibfield  {author} {\bibinfo {author} {\bibfnamefont {M.}~\bibnamefont
  {Kolesik}}, \bibinfo {author} {\bibfnamefont {J.~V.}\ \bibnamefont
  {Moloney}}, \ and\ \bibinfo {author} {\bibfnamefont {M.}~\bibnamefont
  {Mlejnek}},\ }\href {\doibase 10.1103/PhysRevLett.89.283902} {\bibfield
  {journal} {\bibinfo  {journal} {Physical Review Letters}\ }\textbf {\bibinfo
  {volume} {89}},\ \bibinfo {pages} {283902} (\bibinfo {year}
  {2002})}\BibitemShut {NoStop}%
\bibitem [{\citenamefont {Kolesik}\ and\ \citenamefont
  {Moloney}(2004)}]{Kolesik2004}%
  \BibitemOpen
  \bibfield  {author} {\bibinfo {author} {\bibfnamefont {M.}~\bibnamefont
  {Kolesik}}\ and\ \bibinfo {author} {\bibfnamefont {J.~V.}\ \bibnamefont
  {Moloney}},\ }\href {\doibase 10.1103/PhysRevE.70.036604} {\bibfield
  {journal} {\bibinfo  {journal} {Physical Review E}\ }\textbf {\bibinfo
  {volume} {70}},\ \bibinfo {pages} {036604} (\bibinfo {year}
  {2004})}\BibitemShut {NoStop}%
\bibitem [{\citenamefont {Couairon}\ \emph {et~al.}(2011)\citenamefont
  {Couairon}, \citenamefont {Brambilla}, \citenamefont {Corti}, \citenamefont
  {Majus}, \citenamefont {de~J.~Ramírez-Góngora},\ and\ \citenamefont
  {Kolesik}}]{Couairon2011}%
  \BibitemOpen
  \bibfield  {author} {\bibinfo {author} {\bibfnamefont {A.}~\bibnamefont
  {Couairon}}, \bibinfo {author} {\bibfnamefont {E.}~\bibnamefont {Brambilla}},
  \bibinfo {author} {\bibfnamefont {T.}~\bibnamefont {Corti}}, \bibinfo
  {author} {\bibfnamefont {D.}~\bibnamefont {Majus}}, \bibinfo {author}
  {\bibfnamefont {O.}~\bibnamefont {de~J.~Ramírez-Góngora}}, \ and\ \bibinfo
  {author} {\bibfnamefont {M.}~\bibnamefont {Kolesik}},\ }\href {\doibase
  10.1140/epjst/e2011-01503-3} {\bibfield  {journal} {\bibinfo  {journal} {The
  European Physical Journal Special Topics}\ }\textbf {\bibinfo {volume}
  {199}},\ \bibinfo {pages} {5} (\bibinfo {year} {2011})}\BibitemShut {NoStop}%
\bibitem [{\citenamefont {Fonseca}\ \emph {et~al.}(2002)\citenamefont
  {Fonseca}, \citenamefont {Silva}, \citenamefont {Tsung}, \citenamefont
  {Decyk}, \citenamefont {Lu}, \citenamefont {Ren}, \citenamefont {Mori},
  \citenamefont {Deng}, \citenamefont {Lee}, \citenamefont {Katsouleas},\ and\
  \citenamefont {Adam}}]{Fonseca2002a}%
  \BibitemOpen
  \bibfield  {author} {\bibinfo {author} {\bibfnamefont {R.~A.}\ \bibnamefont
  {Fonseca}}, \bibinfo {author} {\bibfnamefont {L.~O.}\ \bibnamefont {Silva}},
  \bibinfo {author} {\bibfnamefont {F.~S.}\ \bibnamefont {Tsung}}, \bibinfo
  {author} {\bibfnamefont {V.~K.}\ \bibnamefont {Decyk}}, \bibinfo {author}
  {\bibfnamefont {W.}~\bibnamefont {Lu}}, \bibinfo {author} {\bibfnamefont
  {C.}~\bibnamefont {Ren}}, \bibinfo {author} {\bibfnamefont {W.~B.}\
  \bibnamefont {Mori}}, \bibinfo {author} {\bibfnamefont {S.}~\bibnamefont
  {Deng}}, \bibinfo {author} {\bibfnamefont {S.}~\bibnamefont {Lee}}, \bibinfo
  {author} {\bibfnamefont {T.}~\bibnamefont {Katsouleas}}, \ and\ \bibinfo
  {author} {\bibfnamefont {J.~C.}\ \bibnamefont {Adam}},\ }\href {\doibase
  10.1007/3-540-47789-6_36} {\bibfield  {journal} {\bibinfo  {journal} {Lecture
  Notes in Computer Science (including subseries Lecture Notes in Artificial
  Intelligence and Lecture Notes in Bioinformatics)}\ }\textbf {\bibinfo
  {volume} {2331 LNCS}},\ \bibinfo {pages} {342} (\bibinfo {year}
  {2002})}\BibitemShut {NoStop}%
\bibitem [{\citenamefont {Fonseca}\ \emph {et~al.}(2013)\citenamefont
  {Fonseca}, \citenamefont {Vieira}, \citenamefont {Fiuza}, \citenamefont
  {Davidson}, \citenamefont {Tsung}, \citenamefont {Mori},\ and\ \citenamefont
  {Silva}}]{Fonseca2013}%
  \BibitemOpen
  \bibfield  {author} {\bibinfo {author} {\bibfnamefont {R.~A.}\ \bibnamefont
  {Fonseca}}, \bibinfo {author} {\bibfnamefont {J.}~\bibnamefont {Vieira}},
  \bibinfo {author} {\bibfnamefont {F.}~\bibnamefont {Fiuza}}, \bibinfo
  {author} {\bibfnamefont {A.}~\bibnamefont {Davidson}}, \bibinfo {author}
  {\bibfnamefont {F.~S.}\ \bibnamefont {Tsung}}, \bibinfo {author}
  {\bibfnamefont {W.~B.}\ \bibnamefont {Mori}}, \ and\ \bibinfo {author}
  {\bibfnamefont {L.~O.}\ \bibnamefont {Silva}},\ }\href {\doibase
  10.1088/0741-3335/55/12/124011} {\bibfield  {journal} {\bibinfo  {journal}
  {Plasma Physics and Controlled Fusion}\ }\textbf {\bibinfo {volume} {55}},\
  \bibinfo {pages} {124011} (\bibinfo {year} {2013})}\BibitemShut {NoStop}%
\bibitem [{\citenamefont {Davidson}\ \emph {et~al.}(2015)\citenamefont
  {Davidson}, \citenamefont {Tableman}, \citenamefont {An}, \citenamefont
  {Tsung}, \citenamefont {Lu}, \citenamefont {Vieira}, \citenamefont {Fonseca},
  \citenamefont {Silva},\ and\ \citenamefont {Mori}}]{Davidson2015}%
  \BibitemOpen
  \bibfield  {author} {\bibinfo {author} {\bibfnamefont {A.}~\bibnamefont
  {Davidson}}, \bibinfo {author} {\bibfnamefont {A.}~\bibnamefont {Tableman}},
  \bibinfo {author} {\bibfnamefont {W.}~\bibnamefont {An}}, \bibinfo {author}
  {\bibfnamefont {F.}~\bibnamefont {Tsung}}, \bibinfo {author} {\bibfnamefont
  {W.}~\bibnamefont {Lu}}, \bibinfo {author} {\bibfnamefont {J.}~\bibnamefont
  {Vieira}}, \bibinfo {author} {\bibfnamefont {R.}~\bibnamefont {Fonseca}},
  \bibinfo {author} {\bibfnamefont {L.}~\bibnamefont {Silva}}, \ and\ \bibinfo
  {author} {\bibfnamefont {W.}~\bibnamefont {Mori}},\ }\href {\doibase
  10.1016/j.jcp.2014.10.064} {\bibfield  {journal} {\bibinfo  {journal}
  {Journal of Computational Physics}\ }\textbf {\bibinfo {volume} {281}},\
  \bibinfo {pages} {1063} (\bibinfo {year} {2015})}\BibitemShut {NoStop}%
\bibitem [{\citenamefont {Li}\ \emph {et~al.}(2021)\citenamefont {Li},
  \citenamefont {Miller}, \citenamefont {Xu}, \citenamefont {Tsung},
  \citenamefont {Decyk}, \citenamefont {An}, \citenamefont {Fonseca},\ and\
  \citenamefont {Mori}}]{Li2021}%
  \BibitemOpen
  \bibfield  {author} {\bibinfo {author} {\bibfnamefont {F.}~\bibnamefont
  {Li}}, \bibinfo {author} {\bibfnamefont {K.~G.}\ \bibnamefont {Miller}},
  \bibinfo {author} {\bibfnamefont {X.}~\bibnamefont {Xu}}, \bibinfo {author}
  {\bibfnamefont {F.~S.}\ \bibnamefont {Tsung}}, \bibinfo {author}
  {\bibfnamefont {V.~K.}\ \bibnamefont {Decyk}}, \bibinfo {author}
  {\bibfnamefont {W.}~\bibnamefont {An}}, \bibinfo {author} {\bibfnamefont
  {R.~A.}\ \bibnamefont {Fonseca}}, \ and\ \bibinfo {author} {\bibfnamefont
  {W.~B.}\ \bibnamefont {Mori}},\ }\href {\doibase 10.1016/j.cpc.2020.107580}
  {\bibfield  {journal} {\bibinfo  {journal} {Computer Physics Communications}\
  }\textbf {\bibinfo {volume} {258}},\ \bibinfo {pages} {107580} (\bibinfo
  {year} {2021})}\BibitemShut {NoStop}%
\bibitem [{\citenamefont {Miller}\ \emph {et~al.}(2023)\citenamefont {Miller},
  \citenamefont {Palastro}, \citenamefont {Shaw}, \citenamefont {Li},
  \citenamefont {Tsung}, \citenamefont {Decyk}, \citenamefont {Joshi},\ and\
  \citenamefont {Mori}}]{Miller2023}%
  \BibitemOpen
  \bibfield  {author} {\bibinfo {author} {\bibfnamefont {K.~G.}\ \bibnamefont
  {Miller}}, \bibinfo {author} {\bibfnamefont {J.~P.}\ \bibnamefont
  {Palastro}}, \bibinfo {author} {\bibfnamefont {J.~L.}\ \bibnamefont {Shaw}},
  \bibinfo {author} {\bibfnamefont {F.}~\bibnamefont {Li}}, \bibinfo {author}
  {\bibfnamefont {F.~S.}\ \bibnamefont {Tsung}}, \bibinfo {author}
  {\bibfnamefont {V.~K.}\ \bibnamefont {Decyk}}, \bibinfo {author}
  {\bibfnamefont {C.}~\bibnamefont {Joshi}}, \ and\ \bibinfo {author}
  {\bibfnamefont {W.~B.}\ \bibnamefont {Mori}},\ }\href {\doibase
  10.1063/5.0152383} {\bibfield  {journal} {\bibinfo  {journal} {Physics of
  Plasmas}\ }\textbf {\bibinfo {volume} {30}} (\bibinfo {year} {2023}),\
  10.1063/5.0152383}\BibitemShut {NoStop}%
\end{thebibliography}%

\end{document}